\documentclass[useAMS,usenatbib]{mn2e}

\def\minus{$\mathrm{-}$}

\newcommand{\erpm}[2]{$\mathrm{_{-#1}^{+#2}}$}
\def\hii{H{\textsc{ii}}}
\def\HI{H{\textsc{i}}}
\newcommand{\timestento}[1]{$\mathrm{\times10^{#1}}$}
\def\microns{$\mathrm{\mu m}$}

\usepackage{subfigure}
\usepackage{graphicx}
\usepackage{wasysym}
\usepackage{stmaryrd}
\usepackage{txfonts}
\usepackage{natbib}
\usepackage{color}
\usepackage{colortbl}

\title[Discs around MYSOs: spectroastrometry of CO at
2.3~$\mu$m]{Probing discs around massive young stellar objects with CO
first overtone emission\thanks{This work is based in part on
observations obtained at the European Southern Observatory Very Large
Telescope under program ID 083.C-0241(A)}\thanks{This paper is also
partly based on observations obtained at the Gemini South Observatory
(program: GS-2007B-Q-214)}}

\author[H. E. Wheelwright et
 al.]{H.E. Wheelwright$^{1}$\thanks{E-mail: pyhew@leeds.ac.uk},
 R.D. Oudmaijer$^{1}$, W.J. de Wit$\mathrm{^{1,2}}$, M.G. Hoare$^{1}$,
 S.L. Lumsden$^{1}$ and \vspace*{1mm}\\
 \LARGE{\rm{J.S. Urquhart$^{3}$}}\\ $^{1}$The School of Physics and
 Astronomy, EC Stoner Building, University of Leeds, Leeds, LS2
 9JT, UK\\ $^{2}$European Organisation for Astronomical Research in
 the Southern Hemisphere, Casilla 19001, Santiago 19, Chile\\
 $^{3}$Australia Telescope National Facility, CSIRO Astronomy and
 Space Science, Sydney, NSW 2052, Australia\\}

\begin{document}

\date{Accepted. Received ; in original form: }

\pagerange{\pageref{firstpage}--\pageref{lastpage}} \pubyear{2010}

\maketitle

\label{firstpage}

\begin{abstract}

We present high resolution ($\mathrm{R\sim50,000}$) spectroastrometry
over the CO $\mathrm{1^{st}}$ overtone bandhead of a sample of seven
intermediate/massive young stellar objects. These are
primarily drawn from the red MSX source (RMS) survey, a systematic
search for young massive stars which has returned a large, well selected
sample of such objects. The mean luminosity of the sample is
approximately 5\timestento{4} $\mathrm{L_{\odot}}$, indicating the
objects typically have a mass of $\sim$15 $\mathrm{M_{\odot}}$. We fit
the observed bandhead profiles with a model of a circumstellar disc,
and find good agreement between the models and observations for all
but one object. We compare the high angular precision
($\mathrm{0.2-0.8}$\timestento{-3} arc-seconds) spectroastrometric
data to the spatial distribution of the emitting material in the
best-fitting models. No spatial signatures of discs are detected,
which is entirely consistent with the properties of the best-fitting
models. {{Therefore, the observations suggest that the CO bandhead
emission of massive young stellar objects originates in small-scale
disks, in agreement with previous work}}. This provides further
evidence that massive stars form via disc accretion, as suggested by
recent simulations.

\end{abstract}


\section{Introduction}

Massive stars ($\mathrm{M_{\star}\ga 8\,M_{\odot}}$) can have a
significant impact upon their environment, despite the fact they are
less numerous than their lower mass counterparts. Massive stars can
dominate their host galaxy's luminosity and inject prodigious amounts
of energy into the interstellar medium (ISM). This injection of energy
can regulate subsequent star formation, and provides a key source of
heating and turbulence in the ISM. Furthermore, the enriched material
injected into the ISM by supernovae accompanying the demise of massive
stars forms a crucial component in subsequent generations of stars and
planets. Therefore, massive stars are important from galactic to
planetary scales \citep{ZinneckerandYorke2007}. However, despite their
importance, our knowledge of how massive stars form is less complete
than in the case of solar mass stars.

\smallskip

There has been considerable theoretical uncertainty over the formation
of massive stars. A young massive star is expected to attain the
luminosity of a main sequence OB star while it is still accreting
material \citep[according to an extrapolation of the standard low mass
star formation scenario of][]{Shu1987}. As a result, it has been
thought that the immense luminosity of massive stars could provide
sufficient radiation pressure to reverse the in-fall of material
\citep{Larson1971,Kahn1974,Wolfire1987}. Consequently, alternative
modes of massive star formation have been proposed, for example
competitive accretion and stellar mergers
\citep{Bonnell1998,Bally2005}. However, recent 3D hydrodynamic
simulations demonstrate that radiation pressure does not prevent disc
accretion forming stars of at least $\mathrm{\sim50\,M_{\odot}}$
\citep{Krumholz2009}. The key detail being that accretion is confined
to an equatorial disc, shielding the accreting material from the brunt
of the radiation, and channeling the radiation pressure into the polar
regions \citep{YorkeandSonnhalter2002,Krumholz2009,Vaidya2009}.

\smallskip

However, relating theoretical models to observations is
challenging. The Kelvin-Helmholtz timescale, the time taken for a
proto-star to convert its potential energy to thermal energy and begin
nuclear fusion, is approximately $\mathrm{10^4}$~years for a massive
star, compared to $\mathrm{\sim10^7}$~years for a star of
$\mathrm{1\,M_{\odot}}$. This short timescale, in conjunction with the
innate rarity of massive stars, makes it difficult to catch a massive
star in the act of forming. In addition, this short time scale means
that massive stars join the main sequence (i.e. begin nuclear fusion
in the core) before their natal material is cleared. Therefore,
massive stars form under many magnitudes of extinction. Finally, as
massive stars are rare, they tend to be further away than nearby sites
of low mass star formation. Consequently, studying massive young
stellar objects (MYSOs) on scales of several au requires
high angular resolution techniques such as spectroastrometry and
optical interferometry \citep[e.g.][]{Davies2010,deWit2010}.

 \smallskip

As a result, direct evidence for accretion discs around MYSOs -- {{which
we define as mid infrared bright stellar objects that have attained
the luminosity of an early OB star but have yet to ionise their
surroundings to form an H{\sc{ii}} region}} -- is sparse. Indirect
evidence that massive stars form by disc accretion is provided by the
discovery of collimated jets and small scale outflows emanating from
MYSOs \citep{Davis2004,Davies2010}. In a few isolated cases, direct
detections of discs and flattened structures around MYSOs have been
made \citep{Shepherd2001,Beltran2005,Patel2005,Okamoto2009}. However,
such studies typically probe dust emission at long wavelengths, and
thus large distances (of order hundreds to thousands of au) from the
central star. Indeed, the rotating structures detected by
\citet{Beltran2005} are too large to be considered circumstellar
discs, as they are unstable and may fragment \citep[see the discussion
in][]{Cesaroni2007}. Therefore, to prove that MYSOs form via disc
accretion requires evidence for small scale, au sized, gaseous discs.

\smallskip

A powerful diagnostic of small-scale discs around young stellar
objects (YSO)s is provided by CO $\mathrm{1^{st}}$ overtone bandhead
emission at 2.3~\microns. Such emission requires high densities
($\mathrm{n\ga10^{10}cm^{-3}}$) and high temperatures
(T$\mathrm{\sim2500K}$), indicative of disc material close to the
stellar surface \citep[e.g.][]{Carr1989}. In addition, such emission
associated with YSOs can be well fit with a model of the bandhead
emission originating in a Keplerian disc, a few au in size
\citep[e.g.][]{Carr1989,Chandler1995}. In particular, \citet{Bik2004}
and \citet{Blum2004} found that the CO bandhead profiles of
intermediate and massive YSOs could be fit with emission from discs
with small, $\approx$ 0.1 au, inner radii. This provides strong
support for the formation of massive stars via disc accretion.
However, the number of objects observed so far is small, and there are
several intermediate mass objects among the observed samples.

\smallskip

In this paper we extend the work of \citet{Bik2004} and
\citet{Blum2004} to a sample of MYSOs drawn from the {{red MSX
(Midcourse Space eXperiment) source}} (RMS) survey catalogue
\citep{RMS}. The RMS is a galaxy wide survey of MYSOs and
the most representative sample of this class of objects. Previous
searches for MYSOs relied on the $IRAS$ Point Source Catalogue
\citep[e.g.][]{Chan1996,Molinari1996,Sridharan2002}. However, due to
the large beam of $IRAS$ (2--5' at 100~\microns), such studies
suffered from considerable source confusion, and were biased away from
the Galactic Plane, where the majority of MYSOs are expected. The RMS,
however, utilises the $MSX$ survey of the galactic plane in the mid
infrared \citep{Price2001}. The $MSX$ survey {{has a resolution of
20\arcsec$\,$ and thus}} offers a factor of 50 improvement in spatial
resolution over $IRAS$, which allows sources to be detected in regions
that are otherwise unresolved. Therefore, the use of the $MSX$ survey
allows a unique and representative sample of MYSOs to be
selected. Specifically, the RMS survey used colour selection criteria
and the $MSX$ and 2MASS catalogues \citep{Egan2003,Cutri2003} to
select an unbiased sample of approximately 2000 potential MYSOs
\citep[see][]{Lumsden2002}. However, this initial sample contains
objects such as planetary nebulae, \hii$\,$regions and low luminosity
YSOs that have a similar appearance to MYSOs in the near and mid
infrared. These contaminant objects have been eliminated via an
extensive multi-wavelength campaign featuring high resolution
(1--2\arcsec) observations in the radio continuum
\citep{Urquhart2007a,Urquhart2009}, the $\mathrm{^{13}CO}$ J=1--0 and
J=2--1 lines \citep{James2007,James2008}, the mid infrared
\citep{Mottram2007} and the near infrared
\citep[e.g.][]{Clarke2006}. {{In total, the RMS
database\footnote{http://www.ast.leeds.ac.uk/RMS/} provides a large
($\sim$500), well-selected sample of mid infrared bright MYSOs.}}

\smallskip 

Here, we exploit the RMS survey to study the accretion characteristics
of a sample of MYSOs. From the low resolution, NIR spectroscopy
undertaken as a part of the classification stage of the survey
\citep[e.g.][]{Clarke2006}, we selected a sub-sample of objects with
CO $\mathrm{1^{st}}$ overtone bandhead emission at 2.3~$\mu$m. We have
observed these objects at high spectral resolution to investigate
whether their CO bandheads are consistent with emission originating in a
small-scale circumstellar disc.

\smallskip

To fully constrain the models, we also require information on the
spatial distribution of the emission. However, the emission region is
expected to be small. At a typical distance of 1kpc, a disc of 1 au
in size subtends an angle of only 1 milli-arcsec
(mas). Spectroastrometry is one of a few approaches that offers the
required, sub-mas, angular precision and high spectral
resolution. Indeed, such an approach has already been shown to be able
to detect and characterise circumstellar discs with a precision of
approximately 0.1 mas (\citealt{Pontoppidanetal2008}, Wheelwright et
al. in prep.). Therefore, we use spectroastrometry to probe the
spatial behaviour of the bandhead emission with high spectral
resolution (which is required to obtain kinematic information).
 
\smallskip

The paper is structured as follows: in Section \ref{Obsanddatared} we
present the sample of MYSOs, details of the observations and the data
reduction processes used. The resultant data are presented in Section
\ref{Results}, alongside the results of fitting a model of CO emission
from a circumstellar disc to the data (the model is described in
Section \ref{model}). Section \ref{disc} presents a discussion of the
results. We conclude the paper in Section \ref{Conclusions}.

\section{Observations and data reduction}
\label{Obsanddatared}

\subsection{Observations}

The sample was selected using the low resolution $K$-band spectroscopy
undertaken as part of the RMS survey (Cooper et al. in prep.) to
identify objects with CO $\mathrm{1^{st}}$ overtone bandhead
emission. In addition to the resulting targets, we included IRAS
08576$-$4334 and M8E in the sample. IRAS 08576$-$4334 is an
intermediate mass YSO known to exhibit CO first
overtone emission \citep{Bik2004}. This object was included in the
sample of MYSOs as its CO emission has been reported to be extended on
scales of tens of au \citep{Grave2007}, making it an appropriate
target to study the spatial distribution of CO emission. M8E is a very
bright MYSO that is thought to possess a circumstellar disc
\citep{Simon1985}. The final sample is presented in Table
\ref{logofobs}, along with details of the observations.

\smallskip

High resolution spectroscopy of the sample at $\mathrm{2.3 \mu m}$ was
obtained using Phoenix at Gemini South \citep{Hinkle2000} and CRIRES
\citep{Kaeufl2004} on UT1 at the VLT. When using Phoenix a slit of
0.34~arcsec was used, which resulted in a spectral resolution of
approximately 50 000, or 6~$\mathrm{km\,s^{-1}}$. The CRIRES
observations were conducted using a slit of 0.4~arcsec, and the same
spectral resolution. Observations were conducted using 4 slit position
angles (PA): $\mathrm{0^{\circ}}$, $\mathrm{90^{\circ}}$,
$\mathrm{180^{\circ}}$ and $\mathrm{270^{\circ}}$, which is a
requirement for accurate spectroastrometry \citep[see][]{Bailey1998a}.
 
\smallskip

The observations with Phoenix were conducted with the K4396 filter,
which has a central wavelength of 2.295~{\microns}, to isolate the
region surrounding the CO $\mathrm{1^{st}}$ overtone bandhead. When
using CRIRES, the K$\mathrm{_s}$ filter was used with the spectral
configuration identified by the reference wavelength
2.2932~\microns. This resulted in the CO $\mathrm{1^{st}}$ overtone
bandhead being located on the third chip, with a few ro-vibrational
lines evident on the fourth (e.g. between $\mathrm{\sim2.30-2.31 \mu
m}$). All observations were conducted using a standard nodding
sequence along the slit to remove the sky background. Where a natural
guide star was available (G332.8256$-$00.5498 and G347.0775$-$00.3927) the
adaptive optic capabilities of the VLT, the Multi-Application
Curvature Adaptive Optics facility, was utilised. All the
observations with Phoenix at Gemini South were conducted with natural
seeing. The resulting Full Width at Half Maximum (FWHM) of the
spectral spatial profiles ranged from 0.27 to 0.75~arcsec, and was
typically 0.5~arcsec.

\smallskip

Telluric standard stars, late B type or early A type dwarfs, were
observed with the identical instrumental setup as the science
observations, and at similar air-masses. These spectra were
subsequently used to correct the science spectra for telluric lines,
and to provide a rough photometric calibration.

\begin{center}
  \begin{table*}
    \begin{center}
      \begin{minipage}{\textwidth}
	\begin{center}
	  \renewcommand{\footnoterule}
	  
	  \caption{A summary of the sample and observations. The
	  bolometric luminosities (Col. 5) are taken from the
	  $\mathrm{RMS\,database}^{\dag}$, which are based on the work
	  of \citet[][]{Mottram2010} and Mottram et al. (2010: in
	  prep.). The kinematic distances (Col. 7) are also taken from
	  the RMS database$\mathrm{^\circ}$, unless taken from the
	  literature as indicated. The masses (Col. 4) are estimated
	  from the bolometric luminosities and main sequence
	  relationships \citep{Martins2005}, unless
	  otherwise stated. The $K$-band magnitudes (Col. 6) are from
	  the $\mathrm{2MASS\,Catalog^{\ddag}}$, unless otherwise
	  stated. The integration time (Col. 8) denotes the total
	  integration time per object, and the precision of the
	  re-binned spectroastrometric data (the average rms of the
	  position spectra) are presented in Col. 9. Finally, we list
	  the dates the objects were observed
	  (Col. 10). \label{logofobs}}

	  \begin{tabular}{p{28.0mm} p{13mm} p{13.5mm} p{5mm} p{6.5mm} p{5mm} p{5mm} p{5mm} p{5mm} l}
	    \hline	  
	     Name & RA & Dec & M & $\mathrm{L_{Bol}}$ & $K$ & $d_{\mathrm{kin}}$ & $\mathrm{Time}$ &$\mathrm{\Delta}$pos &Date\\
	     &   J2000        &    J2000        &  $\mathrm{M_{\odot}}$         &$\mathrm{L_{\odot}}$ & mag & kpc& hours & mas\\
	     \hline
	     \hline
	     \textbf{Gemini South sample:}\\
	     IRAS 08576$-$4334$\mathrm{^\ast}$ & 08 59 25.2 & \minus43 45 46.0 & $\mathrm{6.1^{\diamond}}$ & &9.4$^{\ast}$&$\mathrm{2.2^\star}$ & 4.27 & 0.76&2008: 01/02, 01/10, 01/23\\
	     G287.3716+00.6444 & 10:48:04.6 & \minus58:27:01.0 & 15.4 & 2.9\timestento{4} & 7.5 & 5.6 & 3.20 & 0.25&2008: 01/10,01/21\\	    	      
             M 8E$\mathrm{^\ast}$ & 18 04 53.3 &  \minus24 26 42.3 &   13.5$\mathrm{^{\bigtriangleup}}$ & & 4.4  & $\mathrm{1.9^{\bullet}}$ & 0.27 & 0.47&2007: 10/04\\
	     \textbf{VLT sample:}\\
  	     G308.9176+00.1231 & 13:43:01.6 & \minus62:08:51.3 & 42.6 & 3.9\timestento{5} & 6.4  & 5.3 &0.60 & 0.13 &2009: 04/08, 04/11	\\
             G332.8256$-$00.5498 & 16:20:11.1 & \minus50:53:16.2 & 24.5& 1.3\timestento{5} & 8.9 & $\mathrm{3.8^\Box}$ &2.40 & 0.32&2009: 05/08, 08/19, 09/01, /09/08\\
	     G347.0775$-$00.3927 & 17:12:25.8 & \minus39:55:19.9 & 17.7 &4.3\timestento{4} & 8.5 & 14.9 & 2.80 & 0.15 &2009: 07/17, 07/19, /08/03, /08/15 \\
	     G033.3891+00.1989 & 18:51:33.8 & +00:29:51.0 & 11.8 & 1.3\timestento{4} & 7.2 & 5.5 & 1.22& 0.63 &2009: 05/14, 07/05, 08/02, 08/15, 08/19\\
	     \hline
	  \end{tabular} 
	\end{center}
	\small{$\mathrm{\dag}$: http://www.ast.leeds.ac.uk/RMS/}, \small{$\mathrm{\circ}$: the distances in the RMS database are from \citet{Bronfman1996,James2007,James2008}}, \small{$\mathrm{\ddag}$: \citet{Cutri2003}}, \small{$\mathrm{\star}$: Kinematic distance from the rotation curve of \citet{Brand1993} and a $V_{LSR}$ of $\mathrm{7.5 km\,s^{-1}}$ \citep{Bronfman1996}}, \small{$\mathrm{\bullet}$: \citet{Chini1981}}, \small{$\mathrm{\ast}$: \citet{Bik2006}}, \small{$\mathrm{\bigtriangleup}$: \citet{Linz2009}}, \small{$\mathrm{\diamond}$: based on the photometry and $K$ vs $J-K$ diagram of \citet{Bik2006}, a distance of 2.17~kpc and the data of \citet{Harmanec1988}}, \small{$\mathrm{\Box}$: Near/Far ambiguity, here, we use the smallest possible distance.}\\
      \end{minipage}
    \end{center}
  \end{table*}
\end{center}

\subsection{Data reduction}

The data were reduced in a standard fashion. A master flat frame was
constructed, corrected for dark current and normalised. The individual
exposures were then corrected by the normalised, average flat frame
(the IRAS 08576$-$4334 and G287.3716+00.6444 data were corrected with
a median smoothed flat frame due to a discrepancy between the flat
response and the behaviour of the A and B spectra). Pairs of spectra
defined by the A and B nodding positions were subtracted from
one-another, and the individual intensity spectra were extracted from
the resultant data. Spectroastrometry was conducted by fitting a
Gaussian profile to the spatial profile of the spectrum at each
dispersion pixel. This resulted in a position spectrum; the
photo-centre of the spectrum as a function of wavelength, associated
with each conventional spectrum. The centroid may be determined with
high precision, allowing the centre of the emission to be traced to
within a fraction of a pixel \citep[for example
see][]{Oudmaijer2008,Vanderplas2009,Wheelwright2010}. Visual
inspections were used to ensure the Gaussian profile was a valid
representation of the Point Spread Function. As noted by
e.g. \citet[][]{MTakami2003}, the precision of spectroastrometry
scales linearly with the width and signal to noise ratio (SNR) of the
spatial profile. As a consequence of the high SNR data (typically
200-300), and narrow spatial profiles, we were able to achieve an
angular precision down to approximately 0.2 mas.

 \smallskip

The individual flux spectra at a given PA were combined, as were the
positional spectra, to form an average spectrum for the PA
in question. The intensity spectra were divided by the spectrum of a
telluric standard to correct for telluric absorption lines. Finally,
the intensity spectra at each PA were combined to form the average for
the object in question. Dispersion correction was performed using the
telluric lines in the telluric standard spectra and the high
resolution NIR spectrum of Arcturus presented by
\citet{Hinkle1995}. The dispersion correction typically had an rms of
$\mathrm{0.5~km\,s^{-1}}$, i.e. less than $\frac{1}{10}$
the resolution. The positional spectra at anti-parallel position
angles were combined to form the average North-South and East-West
position spectra as follows: (0-180)/2 and (90-270)/2. This eliminates
artificial signatures \citep[see][]{Bailey1998a}, which do not follow
the slit rotation.

\smallskip

Finally, the object spectra were flux calibrated using the telluric
standards. Dividing the science data by a telluric standard spectrum
allowed the continuum brightness ratio to be determined. The $K$-band
magnitude of the science object was then be estimated from the flux
ratio and the telluric's $K$-band magnitude from 2MASS. The result was
generally within 10 per cent of the published $K$-band magnitude. Therefore,
the objects' $K$-band magnitudes were used to estimate the continuum
flux density. The flux over the bandhead was then determined based on
the observed strength of the CO emission.

\subsection{Extinction determination}

\label{ext_det}

In Section \ref{b_f} we compare the flux densities of the best-fitting
models with the observed values, a consistency check that is often not
performed in similar studies. However, in order to assess the
intrinsic bandhead flux, we must first determine the extinction
towards each object.

\smallskip

To achieve this we use the methodology of \citet{Porter1998}, the NIR
extinction law of \citet{Stead2009} and the low resolution, $H$-band
spectroscopy undertaken as part of the RMS survey. It is important to
note that the general extinction law for the interstellar medium may
not be applicable to regions of high extinction
\citep[e.g. see][]{Moore2005}. However, the flattening of the NIR
extinction law reported by \citet{Moore2005} only becomes apparent at
an extinction of approximately 4 magnitudes in $K$, and in general the
extinction towards MYSOs is less than this
\citep[e.g.][]{Porter1998}. Therefore, the general interstellar
extinction law should be applicable to the sample presented here.

\smallskip

\begin{center}
  \begin{figure*}
    \begin{center}
      \begin{tabular}{l l l}
	\includegraphics[width=50mm]{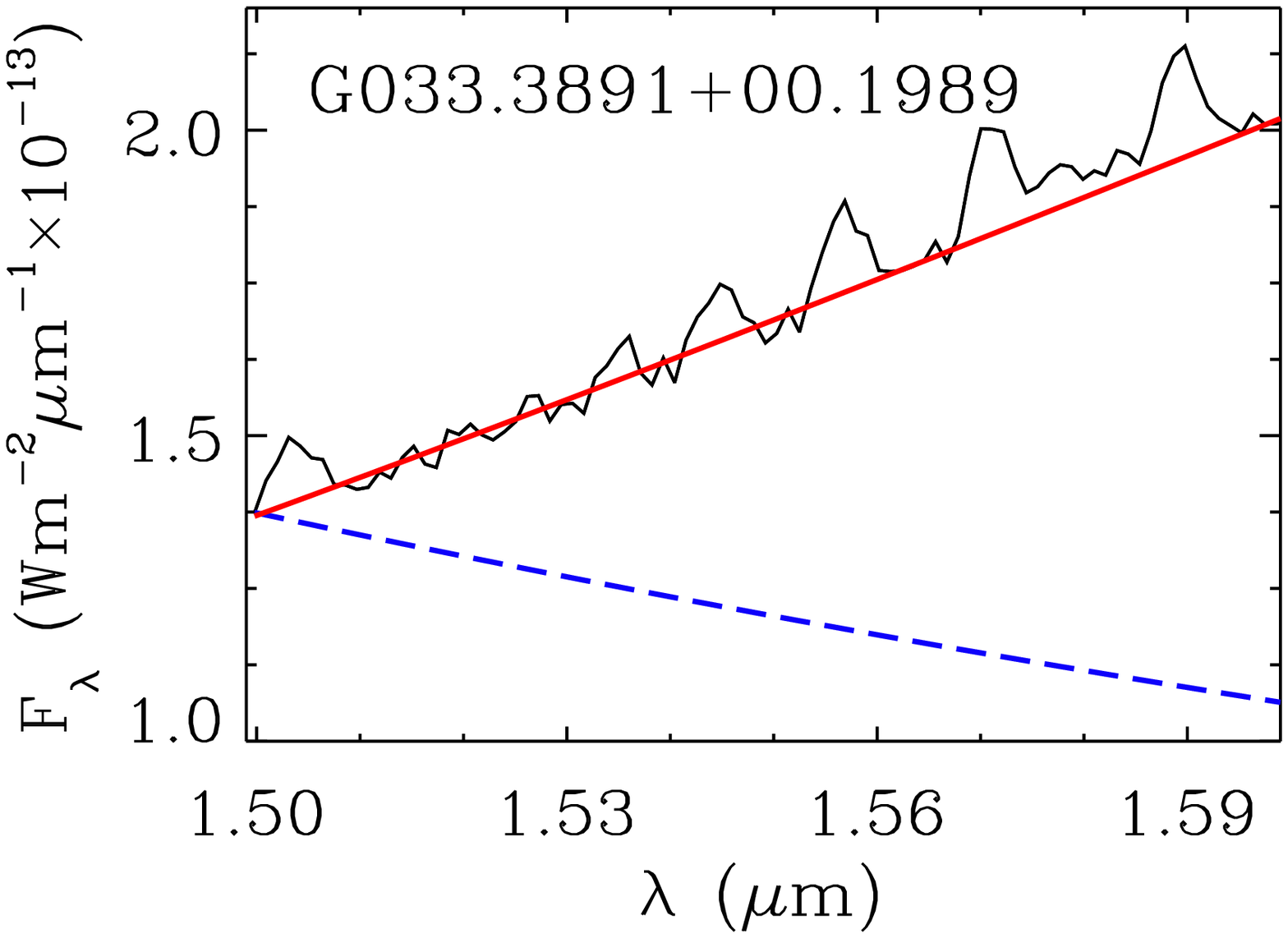} & 
	\includegraphics[width=50mm]{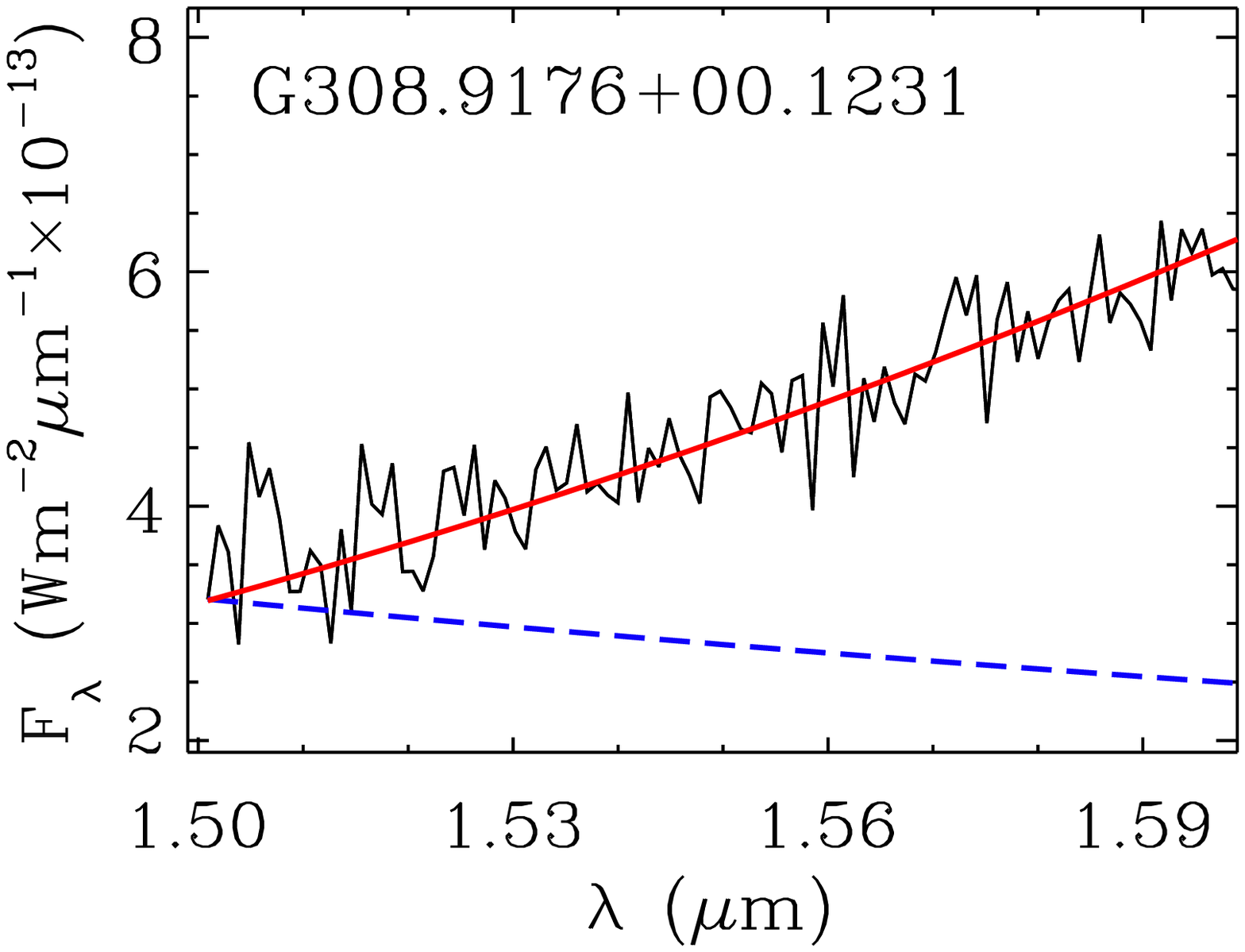} &
	\includegraphics[width=50mm]{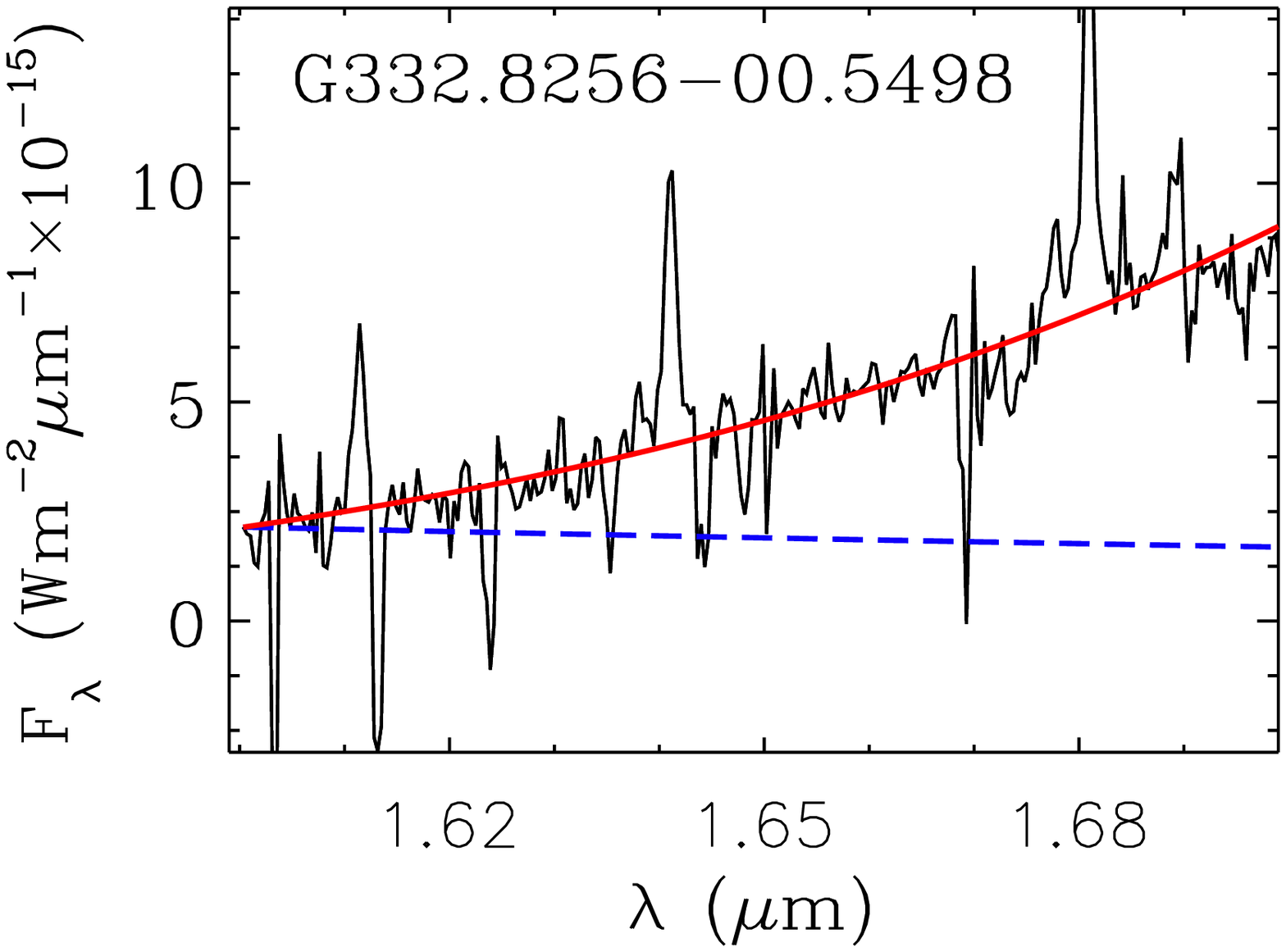} \\
      \end{tabular}
      \begin{tabular}{c c}
	\includegraphics[width=50mm]{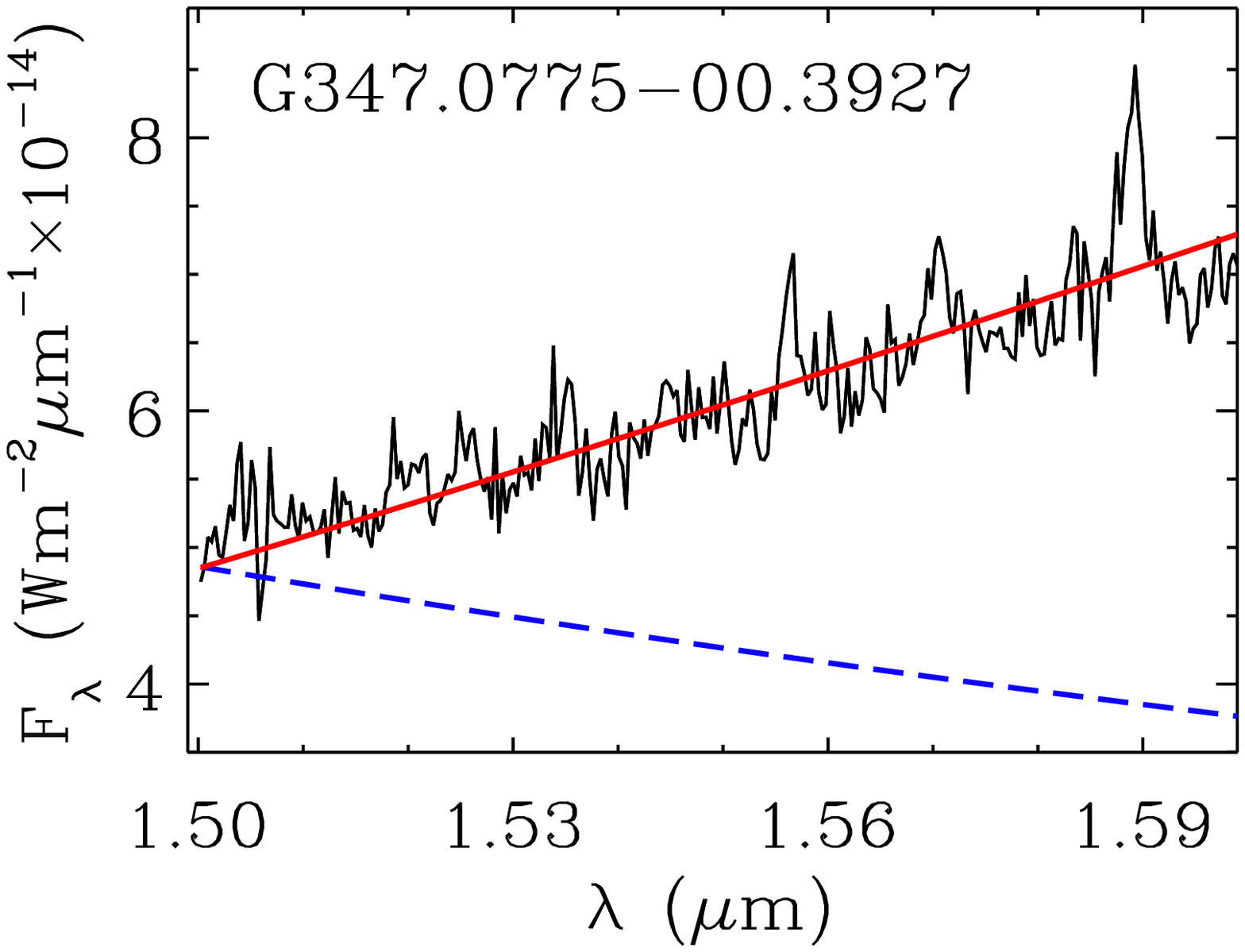} &
	\includegraphics[width=50mm]{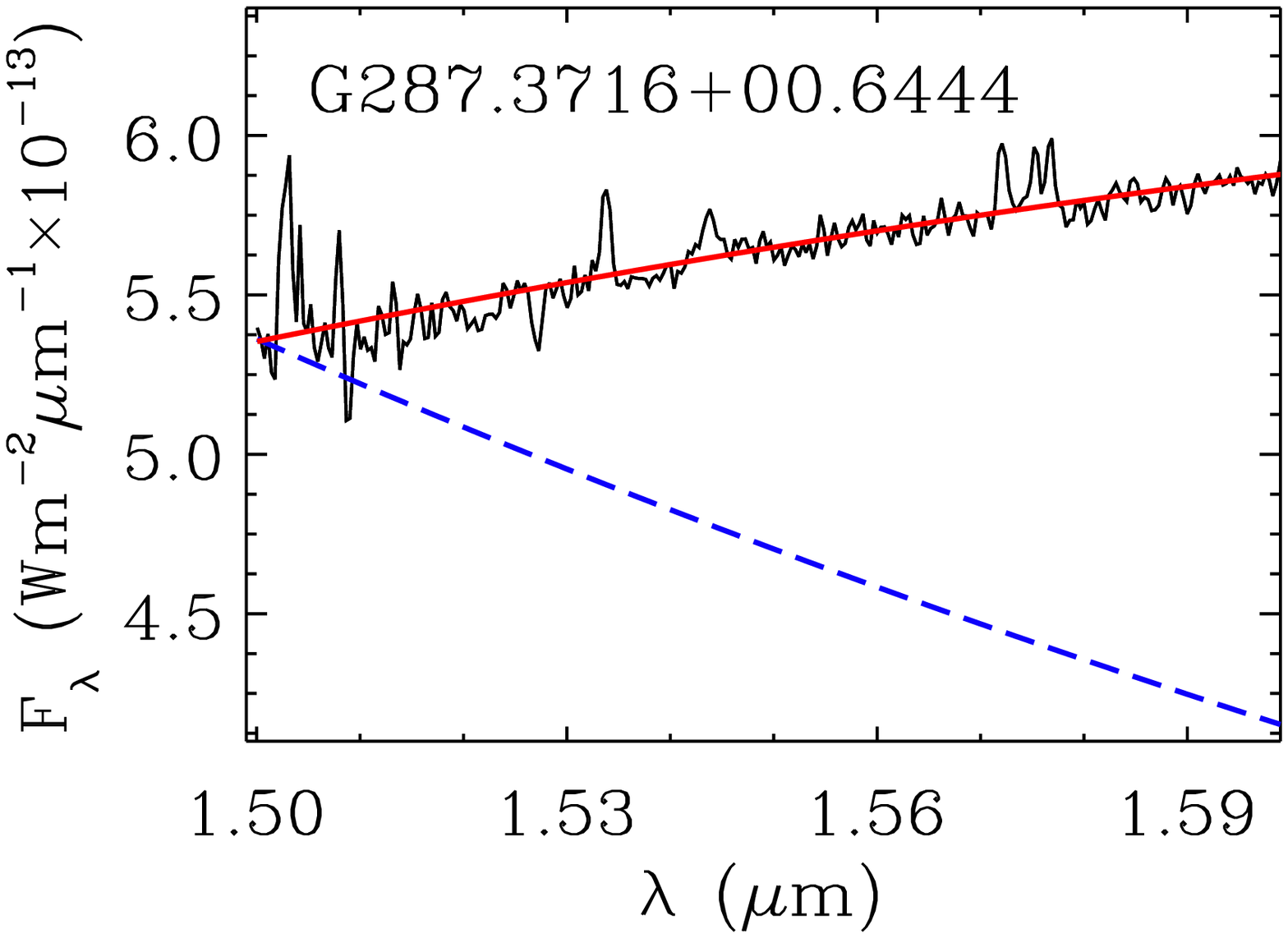} \\
      \end{tabular}
    \end{center}
    
    \caption{\label{red_pic}The RMS sources of which we have obtained
low resolution NIR spectra as part of the follow-up program of
observations for the RMS survey. The \emph{dashed} line
represents the Rayleigh-Jeans tail of a hot star ($F_{\lambda} \propto
\lambda^{-4}$). The \emph{solid} line is this tail reddened to fit the
slope of the $H$-band spectra.}
  \end{figure*}
\end{center}

It is assumed that the science targets are relatively hot, early type
stars. The NIR continuum of a hot star is well approximated by the
Rayleigh-Jeans (RJ) tail \citep{Porter1998}. By reddening the RJ
slope, $\mathrm{F_{\lambda}\propto \lambda^{-4}}$, until it fits the
observed $H$-band continuum, the extinction can be determined. This is
shown in Fig. \ref{red_pic} for the RMS sources for which we have
low resolution NIR spectra. Only the $H$-band spectral continuum is
used, as it is less likely to be contaminated with emission from hot
dust than the $K$-band. Extinction estimates for MYSOs derived from
the continuum in the $H$-band are generally consistent with the
results of other methods, such as using hydrogen recombination line
ratios \citep{Porter1998}. However, extinction values based on the
continuum from the $H$-band and into the $K$-band tend to be
overestimates, due to infrared excess emission from hot circumstellar
dust \citep{Porter1998}. Therefore, we use the wavelength range
$\mathrm{\sim1.5-1.6}$~\microns$\,$ to estimate the extinction towards
the science targets. The determined values of $A_V$ and $A_K$ are
presented in Table \ref{ext}.

\smallskip

The uncertainty in determining the extinction towards an object using
the slope of the its NIR continuum is typically 10--15 per cent
\citep[][based on visually assessing the quality of the
fit]{Porter1998}. We find the uncertainty based on the $\chi^2$
statistic associated with fitting the continuum is also generally 10
percent, or approximately 0.1--0.3 magnitudes. While this uncertainty
is in agreement with previous work there is an important caveat to
consider. This approach assumes the continuum observed is purely
photospheric in origin. If the stellar continuum is contaminated by
hot dust emission, the blue to red slope will become steeper, and the
extinction will be overestimated. Conversely, if continuum is
contaminated by scattered light the extinction will be underestimated
due to a shallower slope.

\smallskip

In general the values of $A_V$ are similar, with most values close to
20--30. However, the derived extinction of $A_V$=70 to
G332.8256$-$00.5498 is significantly higher than the other
values. Contamination of the continuum by dust emission may have
resulted in the extinction being over estimated as the wavelength
range used to determine the extinction was extended to
1.6--1.7~\microns$\,$ (as no flux was observed at
1.5--1.6~\microns). Assessing the presence of dust-excess emission by
SED modelling is beyond the scope of this paper. However, we can use
the ratio of the \HI$\,$ lines in the low resolution spectrum to
assess the extinction independently of the continuum \citep[see
e.g.][]{Lindini1984,Lumsden1996,Moore2005}. Using the ratio of the
B$\gamma$ and Br10 lines and a value of 2.1 for the slope of the NIR
extinction law results in a value of $A_K$=2.4$\mathrm{\pm}$0.3. This
value is consistent with the mean value of the sample, and since it is
not affected by emission from hot dust is likely to be closer to the
correct value than the previous value of 7.9. Therefore, we use this
value of $A_K$ for this object. We note that determining the
extinction towards an object via the ratio of \HI$\,$ recombination
lines is only valid if the intrinsic line ratios can be estimated via
hydrogen recombination models \citep[e.g.][]{Hummer1995}. This
requires that case B of \citet{Baker1938} applies, and in turn limits
this method to \hii$\,$ regions. As G332.8256$-$00.5498 is the only
\hii$\,$ region in the sample this is the only object we can apply
this method to.

\smallskip

These extinction estimates are subject to some uncertainty (of the
order 10 per cent), which translates to a substantial uncertainty in
the de-reddened continuum fluxes (up to approximately 50 per
cent). None the less, the derived fluxes will provide a useful check
on the modelling results.

\begin{center}
  \begin{table*}
    
    \caption{The extinction (Cols 2 \& 3) towards each object
 (Col. 1). The $A_V$ values are determined from the best-fitting $A_K$
 values and the relationship between optical and NIR extinction from
 \citet{Cardelli1989}. In addition, we present the $K$-band continuum
 flux density of each object, based on its $K$-band magnitude
 (Col. 4), and the estimated flux density at the peak of the CO
 $\mathrm{1^{st}}$ overtone bandhead (Col. 5). Finally, we de-redden
 the CO bandhead flux (Col. 6) and compare it to the flux of the
 best-fitting model (Col. 7). The uncertainty in the extinction values
 are estimated from assessing the fit to the continuum with $\chi^2$
 (where the fitting range was specified to avoid lines and artefacts
 in the spectra). The uncertainty in the extinction is then used to
 determine the error in the flux.}
    \label{ext}
    \begin{center}
      \begin{minipage}{\textwidth}
	\centering
   
	\begin{tabular}{l l l l l l l}
	  \hline
           Object & $A_V$ & $A_K$ & $\mathrm{F_{cont}}$ & $\mathrm{F_{CO}}$(observed)  & $\mathrm{F_{CO}}$(de-reddened) & $\mathrm{F_{CO} (model)}$ \\
           &        &       & $\mathrm{Wm^{-2}\mu m^{-1}}$  & $\mathrm{Wm^{-2}\mu m^{-1}}$ & $\mathrm{Wm^{-2}\mu m^{-1}}$ &$\mathrm{Wm^{-2}\mu m^{-1}}$ \\
           \hline
           \hline
           IRAS 08576$-$4334 & 12$^{\dagger}$& 1.4 & 5.8\timestento{-14}& 4.4\timestento{-14} & 1.6\timestento{-13} & 2.8\timestento{-13}\\
           G287.3716+00.6444 & 12 & 1.3$\pm$0.1  & 3.3\timestento{-13} & 2.6\timestento{-14}& 8.6$\pm$0.8\timestento{-14}  & see Section \ref{g287}\\ 
           M8 E & 25$^\ddag$ & 2.8 &  5.8\timestento{-12} & 8.1\timestento{-13} & 1.1\timestento{-11} & 6.2\timestento{-12}\\
           G308.9176+00.1231 & 32& 3.6\erpm{0.5}{0.4} &9.1\timestento{-13} &7.3\timestento{-14}&2.0\erpm{0.7}{0.9}\timestento{-12} & 3.0\timestento{-12}\\
           G033.3891+00.1989 & 22& 2.5$\pm$0.3  & 4.4\timestento{-13}& 5.2\timestento{-14} &5.2\erpm{1.3}{1.6}\timestento{-13}  & 3.7\timestento{-13}\\
           G332.8256$-$00.5498 & 70& 2.4$\pm$0.3  & 9.1\timestento{-14}& 1.9\timestento{-14}  & 1.7\erpm{0.4}{0.6}\timestento{-13} &  3.6\timestento{-13}\\
           G347.0775$-$00.3927 & 23& 2.6$\pm$0.3&1.3\timestento{-13} & 1.3\timestento{-14} & 1.4\erpm{0.3}{0.5}\timestento{-13}& 5.0\timestento{-14}\\
           \hline 
    \end{tabular}

    \small{$\dagger$: \citet{Bik2006}, }
    \small{$\ddag$: \citet{Porter1998}}
      \end{minipage}
    \end{center}
  \end{table*}
\end{center}

\section{Results}
\label{Results}
\label{Discussion}
\subsection{The spectra}

Continuum normalised spectra of the MYSO sample at $\mathrm{2.3\mu}$m
are presented in Fig. \ref{co_spec}. The objects exhibit a range of CO
bandhead profiles. Some display a clear blue `shoulder' at the
bandhead (e.g. G332.8256$-$00.5498), while others exhibit a relatively
sharp rise from the blue continuum to the bandhead (e.g. IRAS
08576$-$4334).  The peak fluxes for the RMS objects are of order 10 per cent
the continuum, much less than that of IRAS 08576$-$4334 ($\mathrm{\sim
80 per cent}$). Bandhead profiles with a prominent blue shoulder are
indicative of emission from a rotating disc
\citep[e.g.][]{Najita1996}. None the less, the profiles with sharper
edges to the bandhead can also be fit with disc models
\citep[see][]{Bik2004}, provided the disc is large or the inclination
is low, thereby minimising the rotational broadening of the bandhead.

\smallskip

We used a simple model to first fit the bandhead profiles. We then
compared the predicted spectroastrometric signatures of the
best-fitting models with the data.

\begin{center}
  \begin{figure*}
    \begin{center}
      \begin{tabular}{l c r}
	
	\includegraphics[width=50mm,height=50mm]{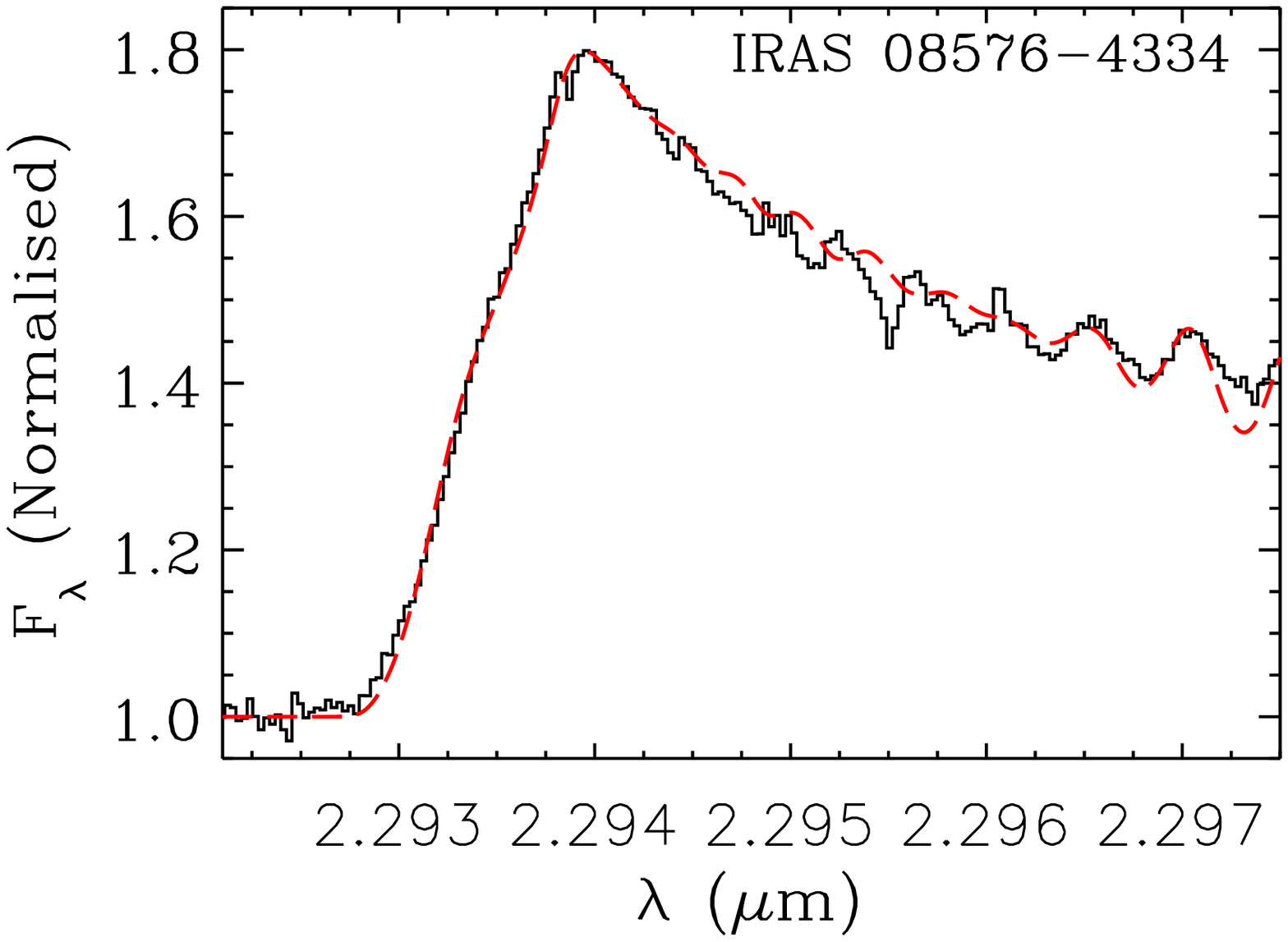} &
	\includegraphics[width=50mm,height=50mm]{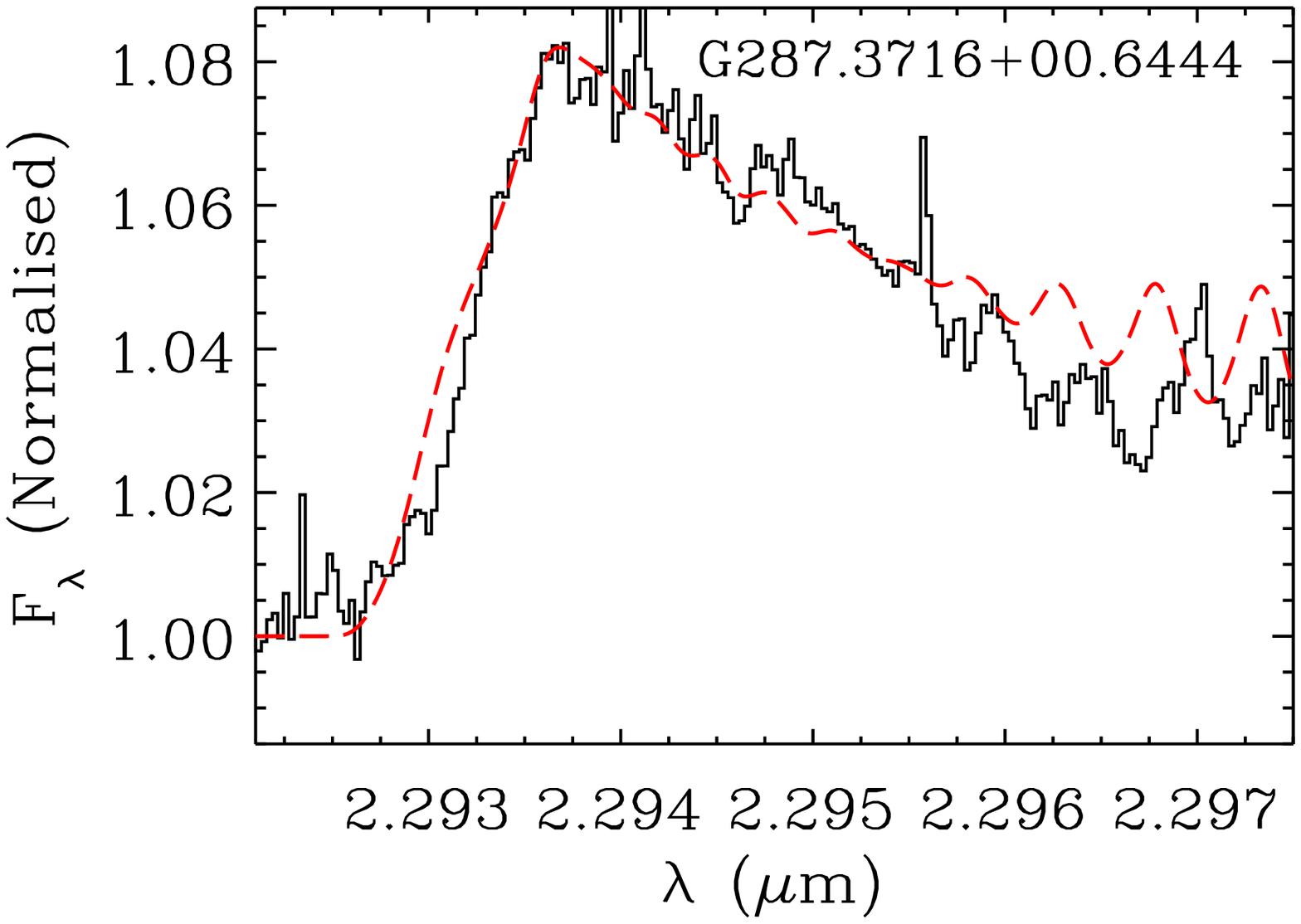}  & 
	\includegraphics[width=50mm,height=50mm]{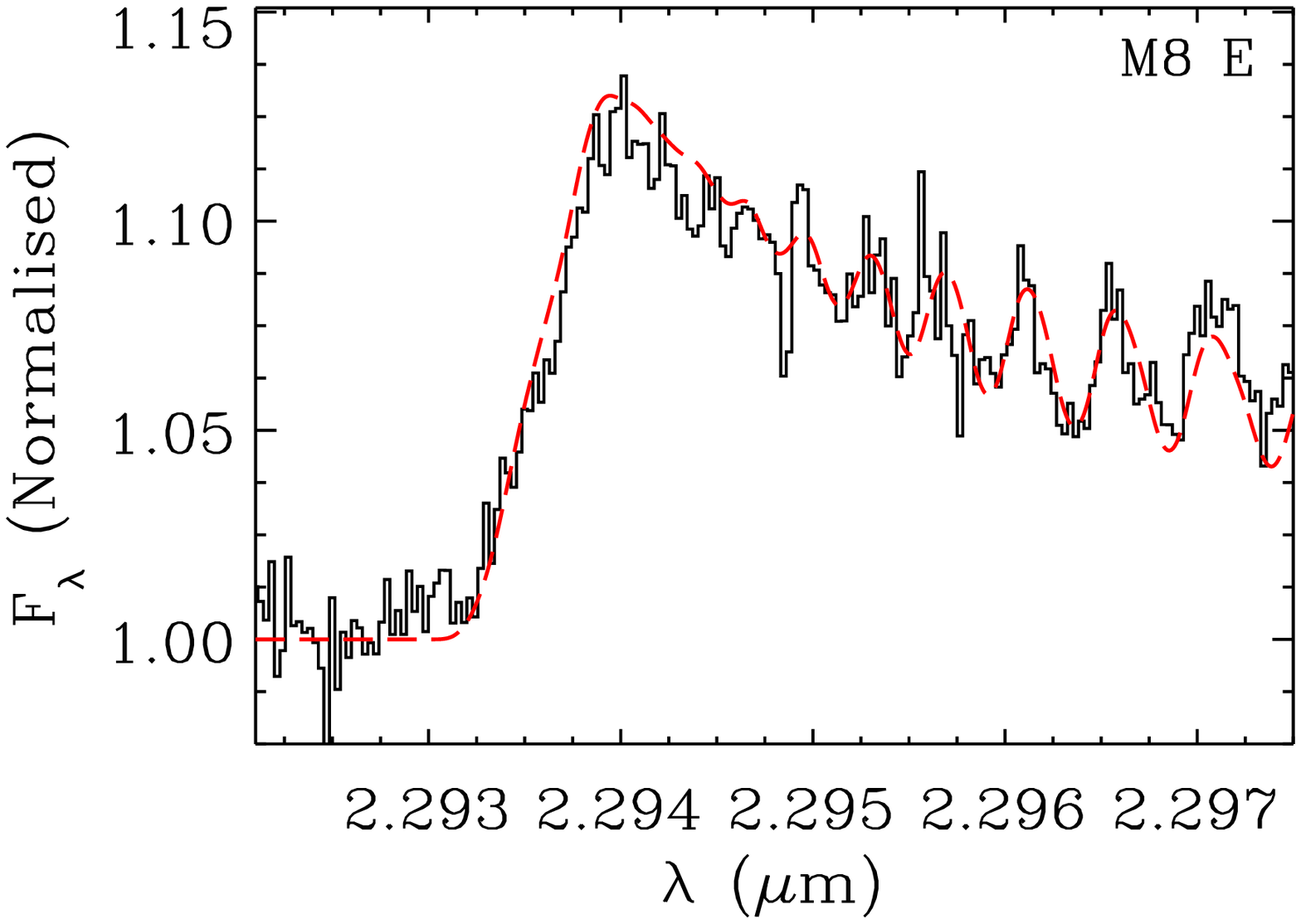} \\

        \includegraphics[width=50mm,height=50mm]{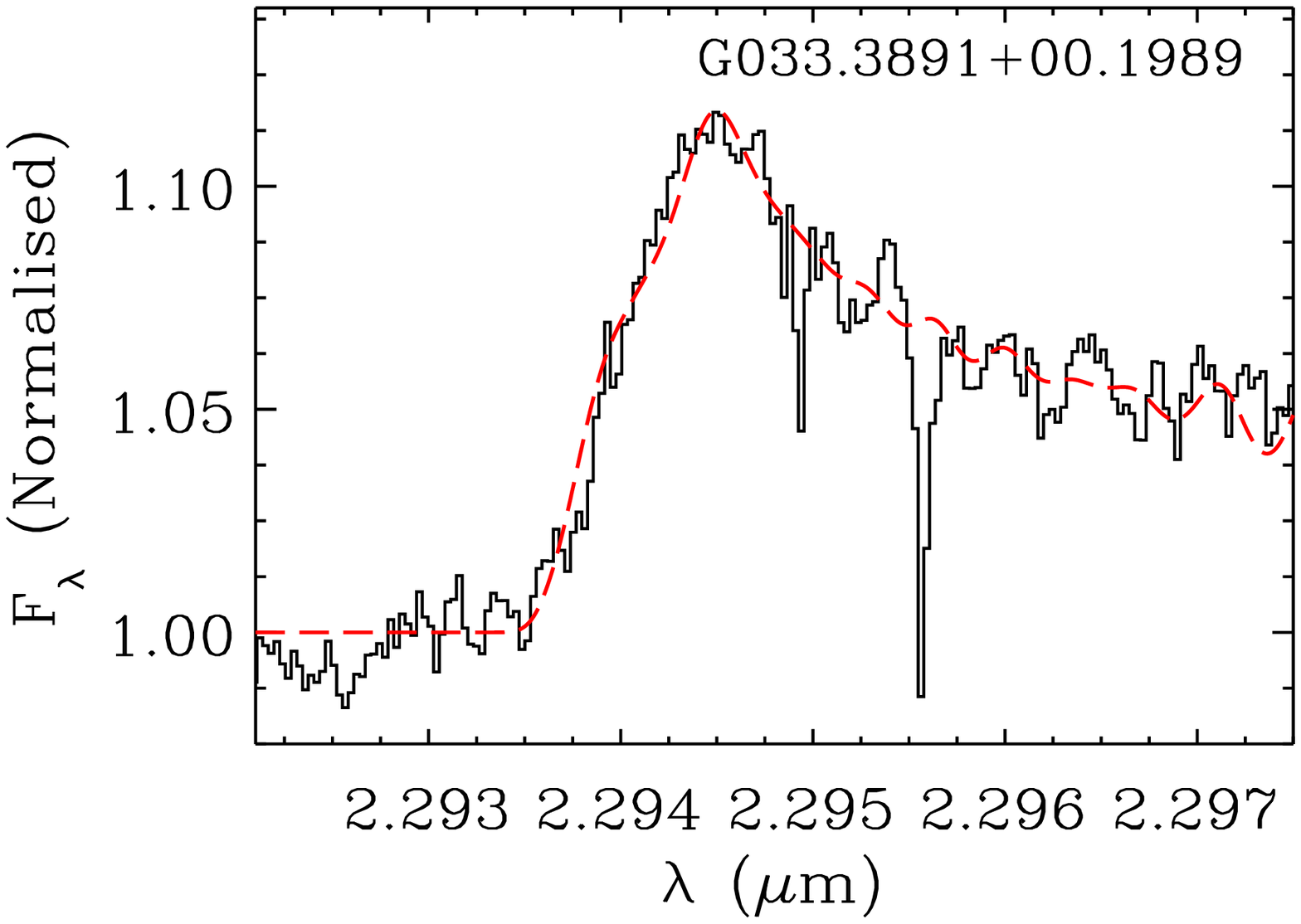}&
	\includegraphics[width=50mm,height=50mm]{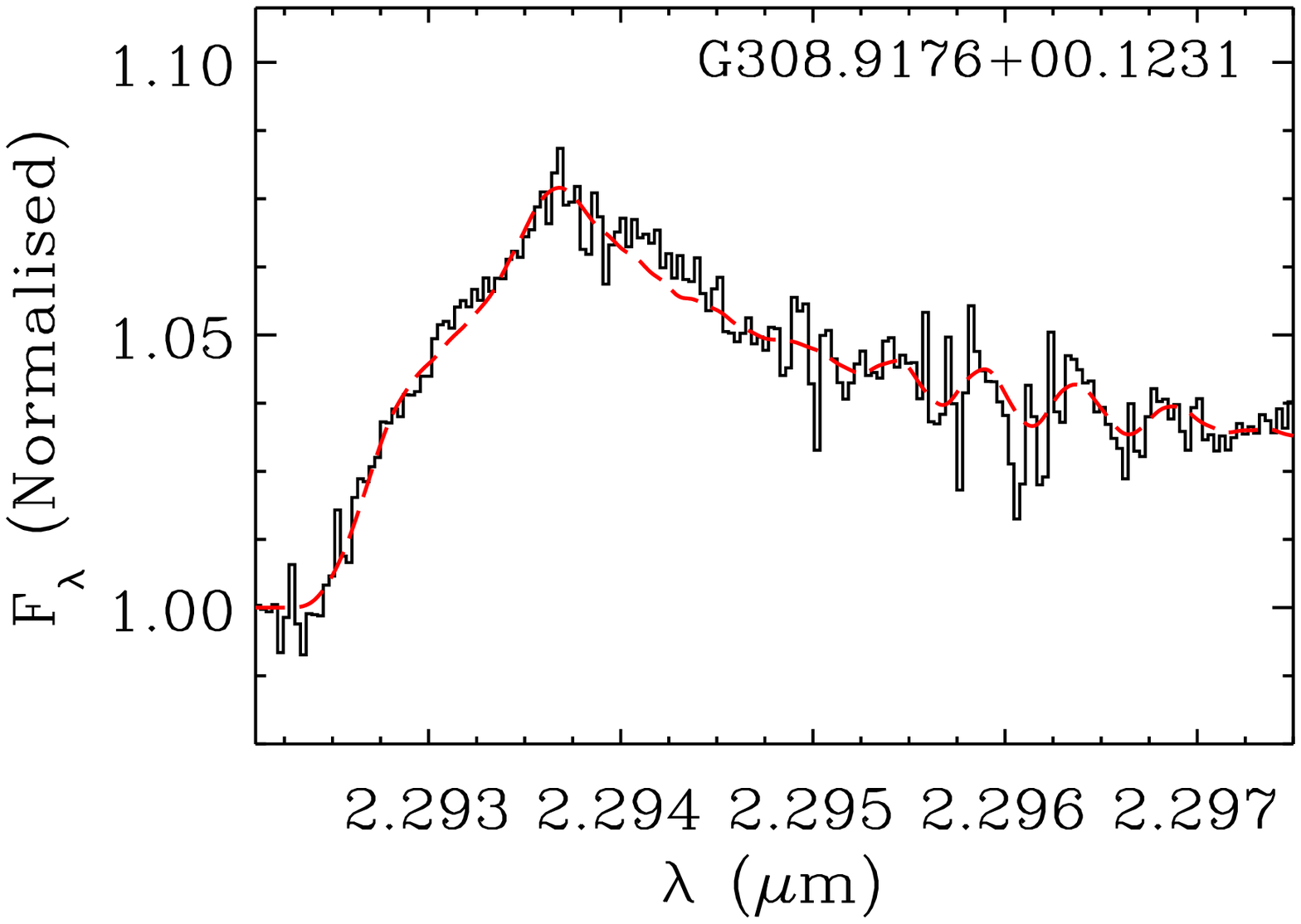} &
	\includegraphics[width=50mm,height=50mm]{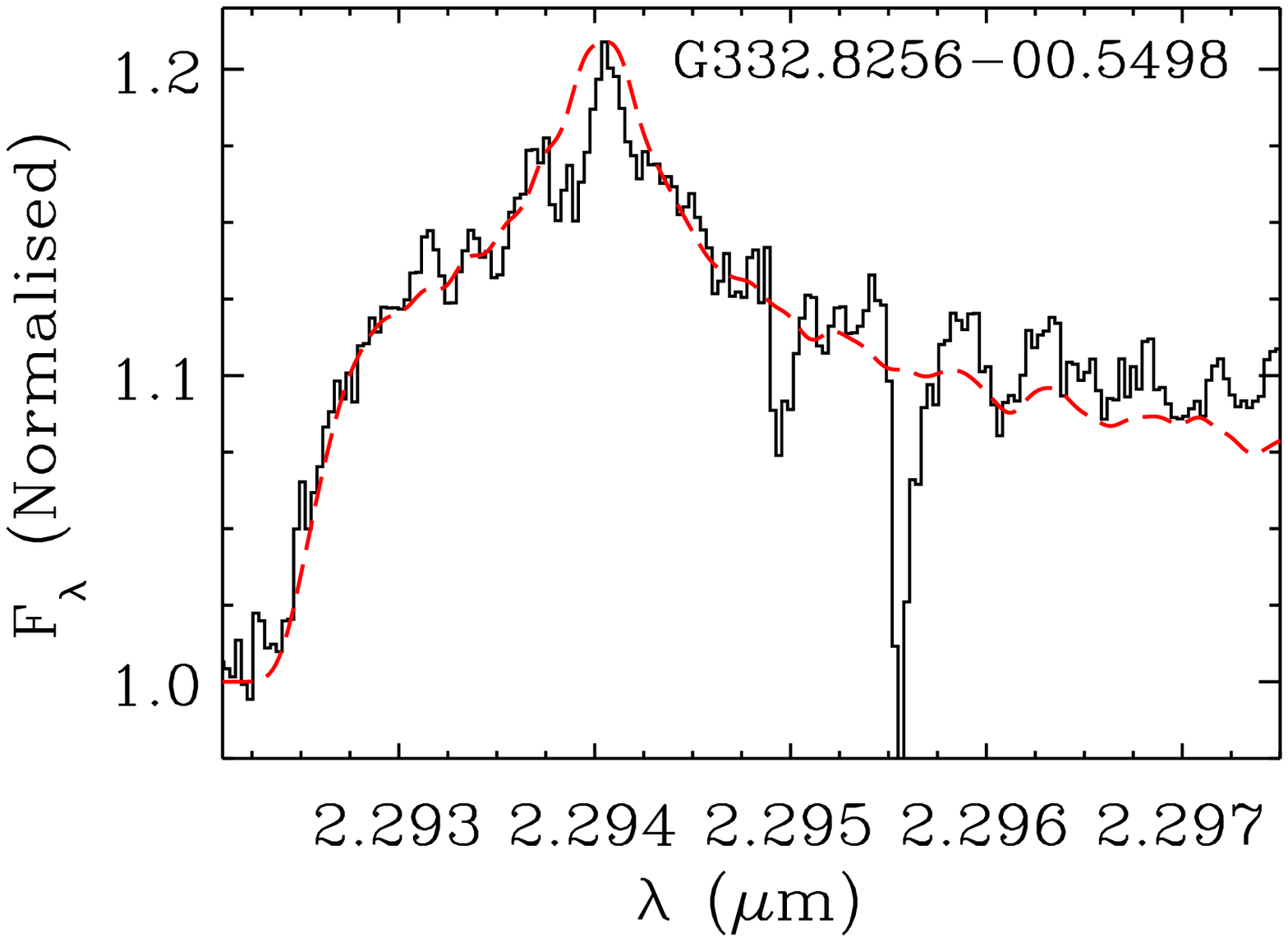} \\
      \end{tabular}
      \begin{center}
	\begin{tabular}{c}
	  \includegraphics[width=50mm,height=50mm]{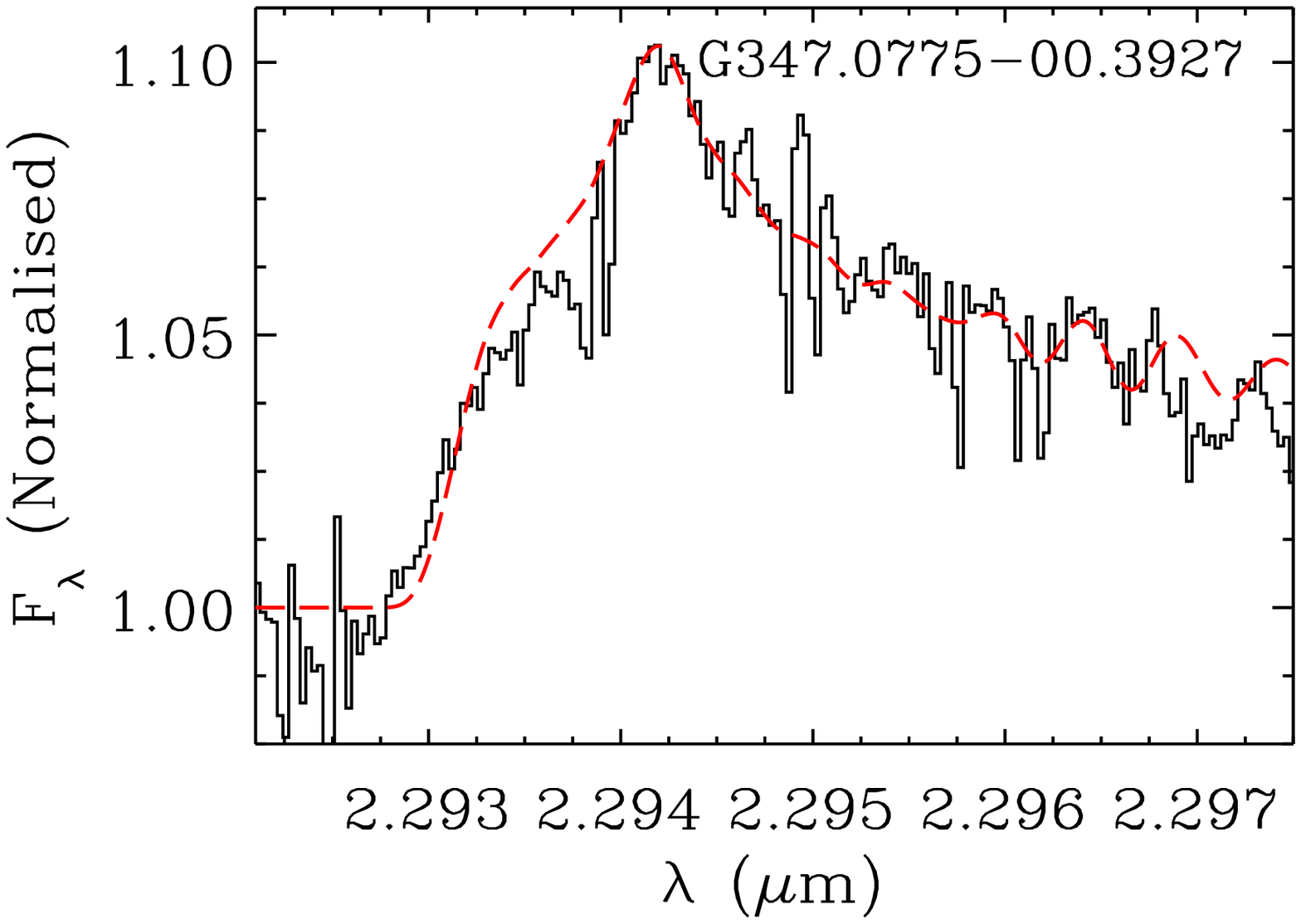}  \\
	\end{tabular}
      \end{center}
      \caption{The spectra of the sample around the CO
	 $\mathrm{1^{st}}$ overtone bandhead. The spectra have been
	 divided by standard spectra, normalised, and re-binned by a
	 factor of 3. \label{co_spec}The \emph{dashed} line represents
	 the best-fitting model, which is described in Section
	 \ref{model}. The absorption features present in the spectra
	 of G033.3891+00.1989 and G332.8256-00.5498 are residual
	 telluric effects.}
      \label{co_spec}
    \end{center}
  \end{figure*}
\end{center}

\begin{center}
  \begin{figure*}
    \begin{center}
      \begin{tabular}{l c r}
	
	\includegraphics[width=50mm,height=50mm]{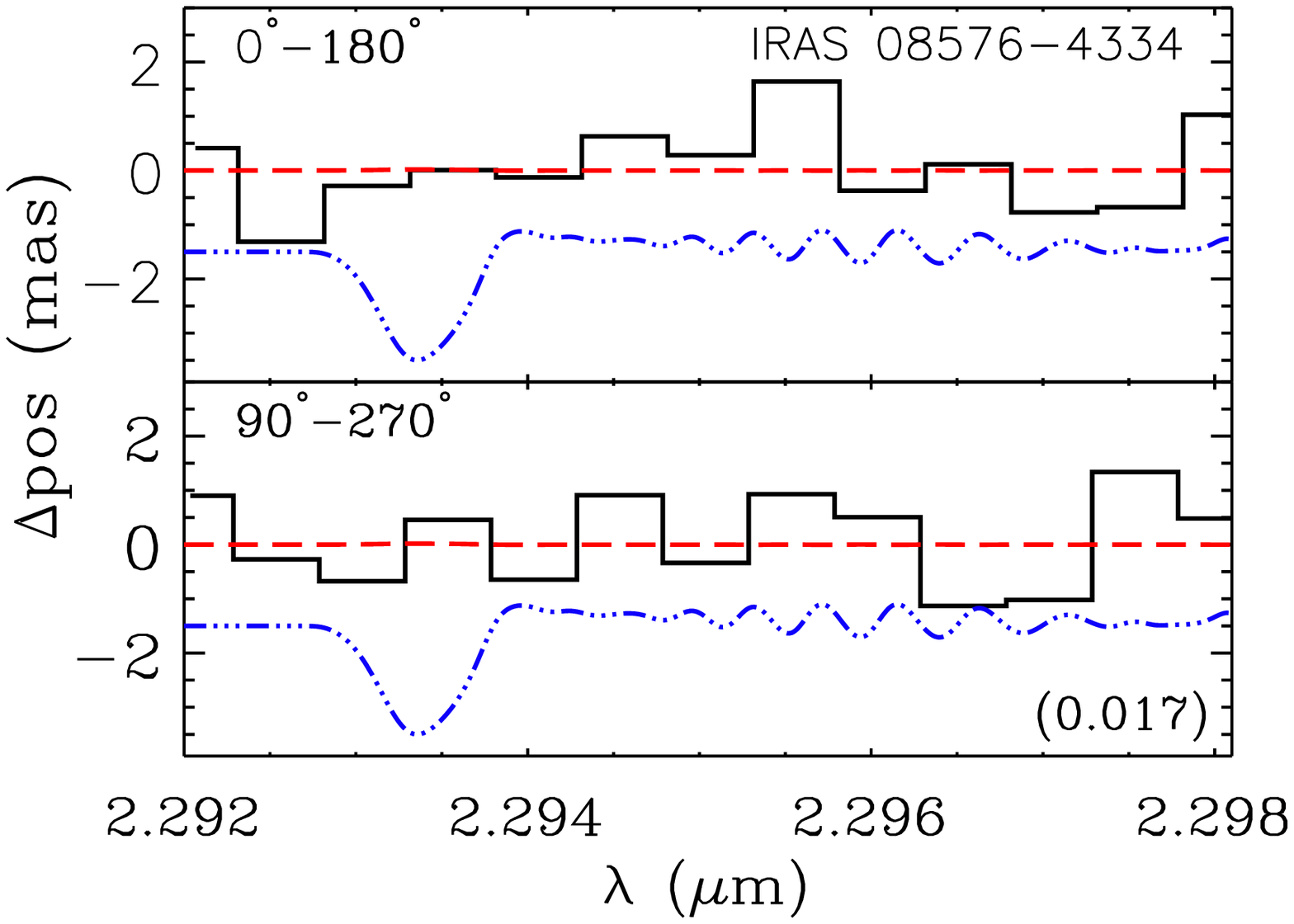} &
	\includegraphics[width=50mm,height=50mm]{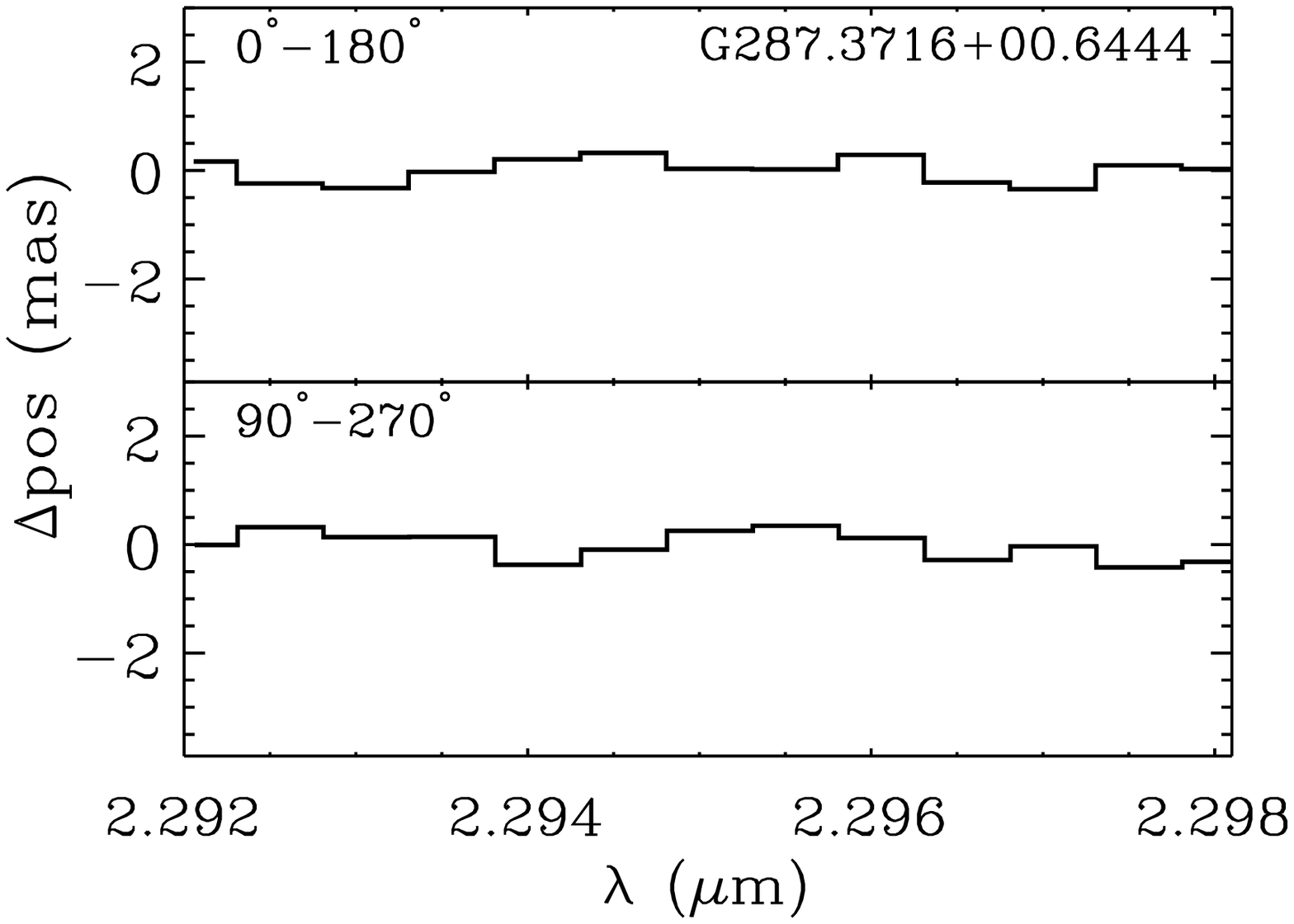}  & 
	\includegraphics[width=50mm,height=50mm]{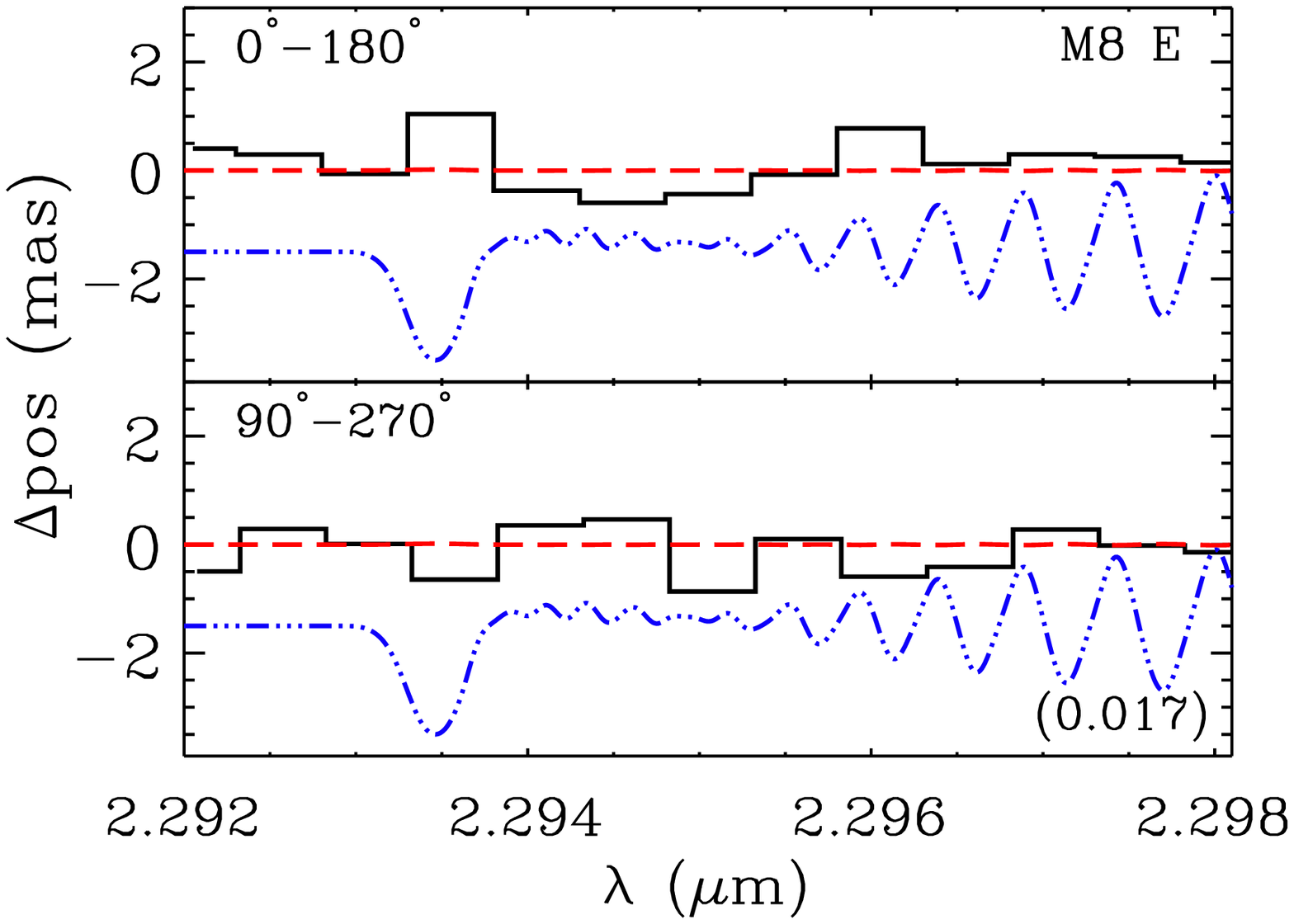} \\
        \includegraphics[width=50mm,height=50mm]{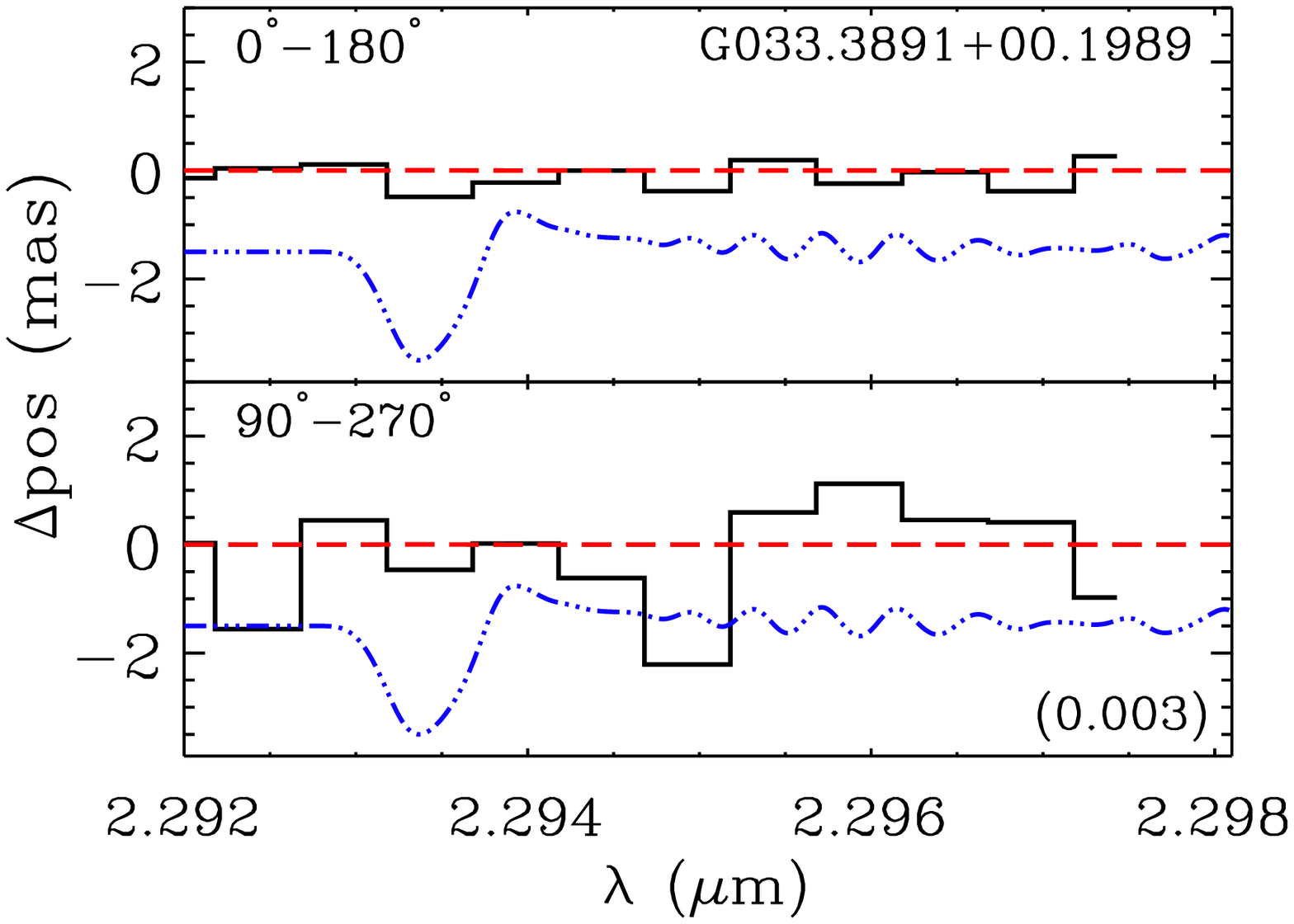}&
	\includegraphics[width=50mm,height=50mm]{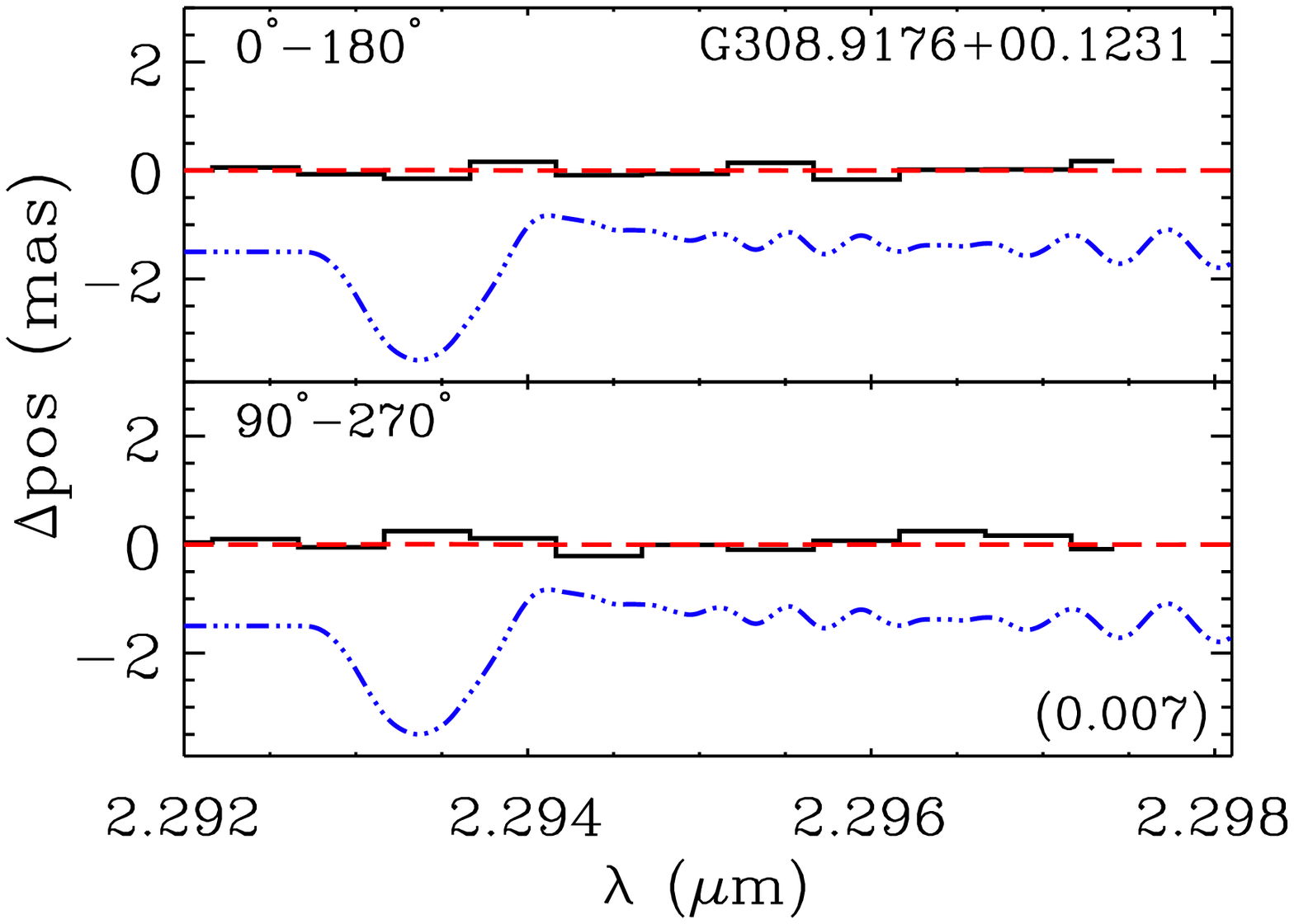} &
	\includegraphics[width=50mm,height=50mm]{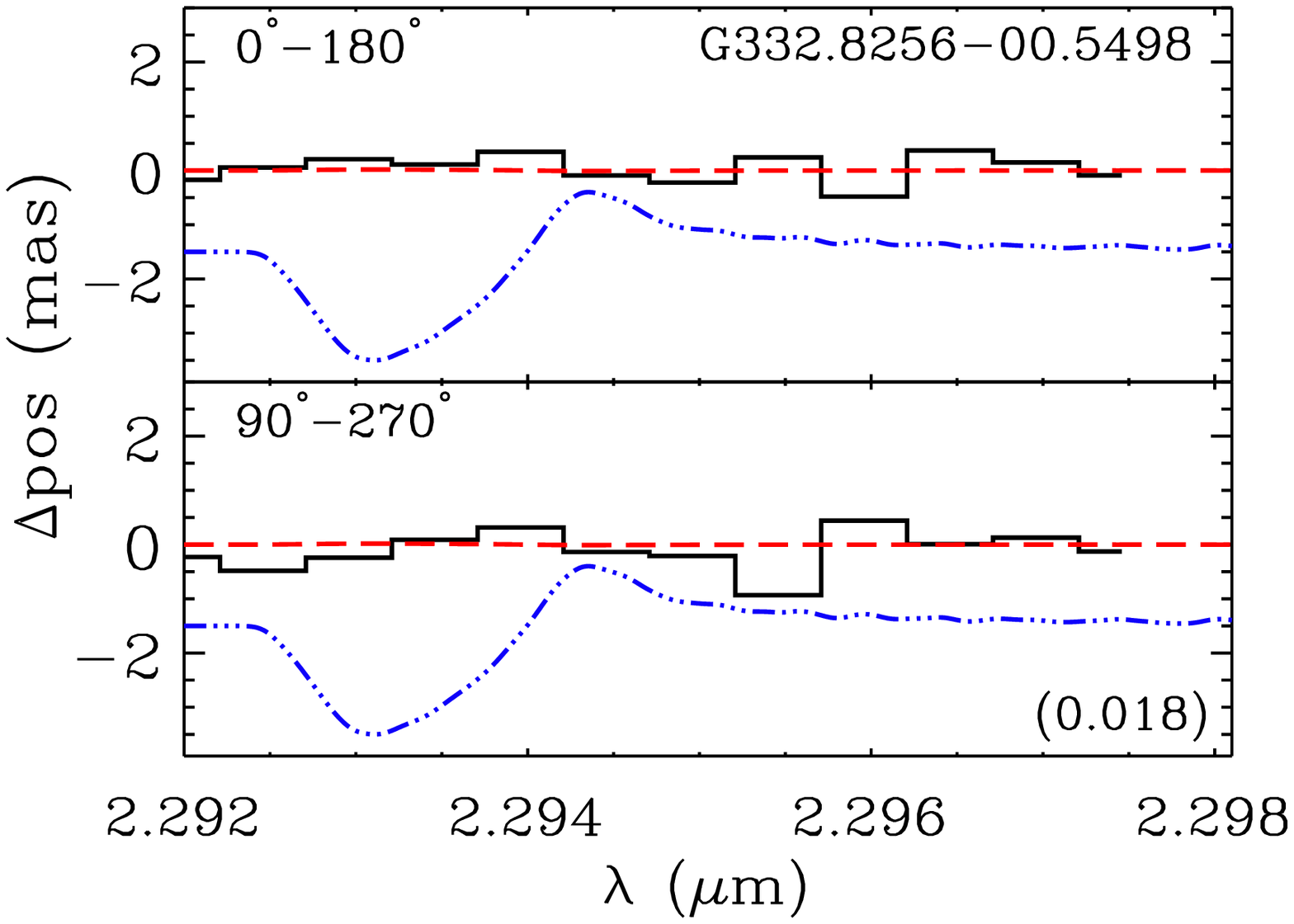} \\
      \end{tabular}
      \begin{center}
	\begin{tabular}{c}
	  \includegraphics[width=50mm,height=50mm]{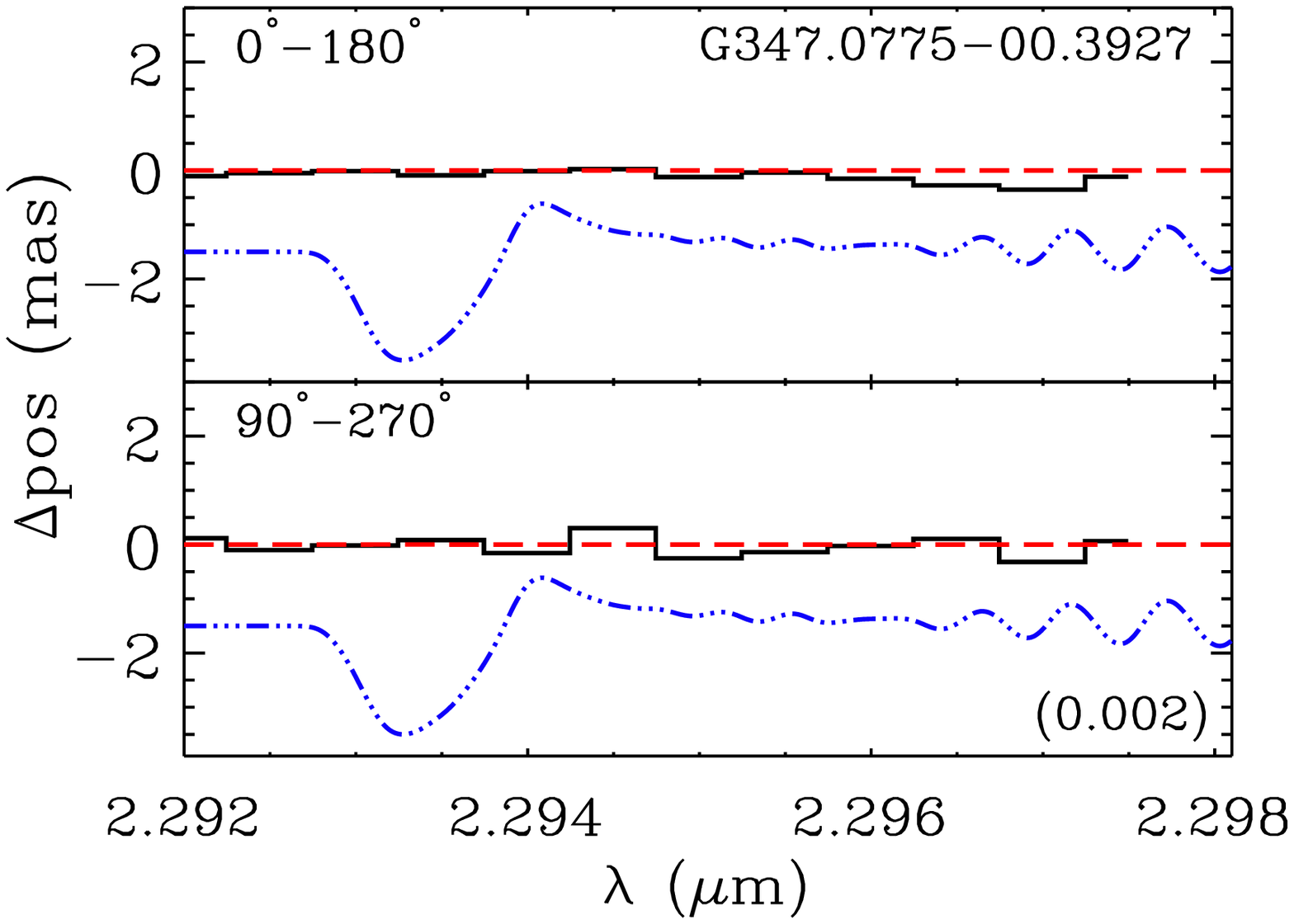}  \\
	\end{tabular}
      \end{center}
      \caption{The spectroastrometric signatures of the sample over
	the CO $\mathrm{1^{st}}$ overtone bandhead emission, shown
	normalised to the continuum position. The data, the
	\emph{solid} line, have been re-binned with a bin size of
	65~$\mathrm{km\,s^{-1}}$ (approximately 11 times the resolution element). We also present the predicted signatures
	of the best-fitting model (see Section \ref{b_f}) for each object,
	bar G287.3716+00.6444 (see Section \ref{g287}), to illustrate
	the signature of a rotating disc. The model signature is shown
	twice (at the un-binned spectral resolution): 1. with its
	predicted amplitude (\emph{dashed} line), and 2. with an
	amplitude of 2~mas in the negative direction to enhance its
	visibility (offset \emph{dot-dashed} line). The maximum
	predicted signature is presented in the \emph{lower right}
	corner of each plot (in mas).}
      \label{spec_ast_obs}
    \end{center}
  \end{figure*}
\end{center}

\subsection{The model}
\label{model}

The model of CO $\mathrm{1^{st}}$ overtone bandhead emission is based
on \citet{Kraus2000}. Initially, a simple, geometrically flat disc is
constructed, within which the excitation temperature and surface
number density decrease with increasing radius. The decrease in
temperature and surface number density are treated as power laws. The
temperature decreases with $R\mathrm{^{-0.75}}$ and the surface number
density falls off with $R\mathrm{^{-1.5}}$, in line with standard
accretion disc theory \citep{Pringle1981}, and following
\citet{Kraus2000}, who also modelled CO emission from a circumstellar
disc. As the disc model is very simple and does not incorporate
accretion the temperature of the disc is determined as follows:
$T(R)=T_{\rm{eff}}\times(\frac{R}{R_{\star}})^{-0.75}$, where
$T_{\rm{eff}}$ is set based on the luminosity of the star (assuming it
to be on the main sequence). The disc is split into radial and
azimuthal cells. If the temperature in a cell is greater than 5000~K
{{(the assumed destruction temperature of CO)}}, the flux from the cell is
set to zero. If the outer radius of the disc is such that the outer
cell temperatures are less than 1000~K the disc is shrunk until the
temperature at the edge is 1000~K. Therefore, while the radii were
varied during the fitting process, they are tied to the temperature
structure of the disc. The CO emission of each cell is calculated
according to the methodology of \citet{Kraus2000}, which is briefly
described in the following.

\smallskip

The population of the CO rotational levels {{(up to J=100)}} for
the 2-0 vibrational transition in each cell is determined assuming
local thermodynamic equilibrium. Then, the absorption coefficient is
determined from the populations of the possible energy levels, and the
transition probabilities of the respective lines. Assuming the
absorption coefficient is constant along the line of sight, the
optical depth is given by the product of the absorption coefficient
per CO molecule and the CO column density. The column density is given
by the surface number density, as we use a thin disc. The Dunham
coefficients required to calculate the CO energy levels were taken
from \citet{Farrenq1991}, and the Einstein coefficients of each
ro-vibrational transition were taken from \citet{Chandra1996}. The
intrinsic line profile is assumed to be Gaussian, with the line-width
being a free parameter. Once the individual cell spectra are
calculated, they are summed to create the total spectrum. This is then
convolved with a Gaussian profile to match the spectral resolution of
the observations.

 \smallskip

To determine the best-fitting model, the downhill simplex algorithm was
used. This algorithm is supplied with the Interactive Data Language
distribution (IDL), and is based on the {\textsc{amoeba}} routine. The
free parameters are: the inner and outer radius of the disc, the
inclination of the disc, the surface number density at the inner edge
of the disc and the line width within the disc. Once the best-fitting
parameters have been determined the spectroastrometric signature of
the best-fitting model is predicted. This is done by mapping the flux from
each cell onto an array with orthogonal spatial and dispersion axes
that represents a long-slit spectrum. The array is then convolved
with a Gaussian profile in the spatial direction, to represent the
seeing conditions. The synthetic signature is then generated by
fitting a Gaussian profile to the spatial distribution at each
dispersion pixel.

\smallskip

The stellar mass is not a free parameter and is set to that listed in
Table \ref{logofobs}, which is generally determined from the
luminosity of the source and main sequence relationships
\citep{Martins2005}. However, the luminosities of IRAS
08576$-$4334 and M8E are not known. Therefore, the mass of IRAS
08576$-$4334 was determined from its position in the $K$ vs $J-K$
diagram of \citet{Bik2006}. We revise the distance to IRAS
08576$-$4334, and find it is still located in the mid- to early B area
of the $K$ vs $J-K$ diagram, and hence assume it has the mass of a B3
type star. The mass of M8E was taken from \citet{Linz2009}, and is the
mass of the best-fitting model from the grid of models by
\citet{Robitaille2007}.

\smallskip

Using the luminosity of a MYSO to determine its mass is subject to
some uncertainty as the observed luminosity may be due in part to
accretion, resulting in an overestimate of the stellar mass. To
evaluate the possible effect this may have, we consulted the models of
accreting, massive protostars presented by
\citet{Hosokawa2009}. During the early adiabatic accretion phase, the
accretion luminosity of a massive protostar is greater than its
intrinsic luminosity. However, by the time a massive protostar enters
its main sequence accretion phase the intrinsic luminosity is greater
than the accretion luminosity. While the resultant total luminosity is
slightly greater than the zero age main sequence luminosity the two
are not different by an order of magnitude. Given the high luminosity
of the sample objects, it is likely they are in the Kelvin-Helmholtz
contraction or main sequence accretion phases. Therefore, it is
unlikely we significantly overestimate their luminosity. Furthermore,
given that the objects are most likely in their main sequence
accretion phase, the main sequence relationship between mass and
luminosity should be applicable. Therefore, it is surmised that using
main sequence relationships is sufficient for the purposes of this
paper.

\subsection{Model fits and results}

\label{b_f}

Over the wavelength range at our disposal few individual
ro-vibrational lines are visible. The fitting process is therefore
dominated by the shape of the CO bandhead peak and shoulder. The
appearance of the shoulder is set by the rotational profile of the
individual lines, and is thus dependent on the inclination and the
inner and outer disc radii (as the stellar mass is not
varied). However, the temperature of the disc is also a function of
distance from the central star. In general few individual lines are
evident, indicative of relatively high temperatures -- and/or high
rotational broadening. As a result, the inner disc radii are generally
small, of the order 0.1~au. Changing the outer radii has little effect
on the quality of the fit as the surface number density, and thus flux,
fall off steeply. As a consequence, it is the inclination that dominates
the fit to the bandhead profile. The object with the most prominent
blue shoulder is fit with the largest inclination (G332.8256$-$00.5498:
$\mathrm{\sim42^\circ}$), while the objects with the steepest slopes
from the blue continuum to the bandhead peak (IRAS 08576$-$4334 and M8E)
are fit with the lowest inclinations ($\mathrm{\sim18^{\circ}}$ \&
$\mathrm{16^{\circ}}$).

\smallskip

The best fit parameters are presented in Table \ref{fit}, and the
resultant bandhead profiles are plotted with the data in Fig.
\ref{co_spec}.  As can be seen from the reduced $\mathrm{\chi^2}$
values (the mean is 1.68) the model of CO emission originating from a
circumstellar disc generally provides a good fit to the data. The sole
exception to this is G287.3716+00.6444, which is discussed in more detail in
Section \ref{g287}.

\smallskip

We compare the predicted bandhead fluxes with the un-reddened observed
fluxes as a consistency check. As mentioned in Section \ref{ext_det},
these un-reddened fluxes are inevitably subject to considerable
uncertainties, yet still serve as an important validity test. We list
both model and source fluxes in Table \ref{ext}. These are similar to
within a factor of 2--3, and generally consistent within the
uncertainty from the extinction estimate (neglecting the addition
error due to uncertainties in the kinematic distances). Therefore, the
observed and model fluxes are essentially consistent. This provides a
further confirmation that disc models provide a good fit to the data.

\begin{center}
  \begin{table*}
 
    \begin{center}
      \begin{minipage}{\textwidth}
	\begin{center}
	  \renewcommand{\footnoterule}
	 
	  \caption{The parameters of the best-fitting models. The mass is determined
	  from the luminosity presented in Table \ref{logofobs}. The
	  uncertainties in $i$, $NCO_{\rm{in}}$ and $\Delta v$ are
	  determined by holding all other parameters constant and
	  determining the change required to increase the reduced
	  $\chi^2$ by 1.0.\label{fit}}
	  \begin{tabular}{l l l l l l l l} 
	    \hline
	     Name & Mass & Inclination & ${R_{\rm{in}}^{\ddag}}$ & ${R_{\rm{out}}^{\ast}}$ & $NCO_{\rm{in}}$ & $\mathrm{\Delta}v$ & $\chi^2$\\
	     & ($\mathrm{M_{\odot}}$) & ($\mathrm{^{\circ}}$) & (au) & (au) & ($\mathrm{cm^{-2}}$) & ($\mathrm{km\,s^{-1}}$) & \\
	    \hline
	    \hline
	    IRAS 08576$-$4334   & 6.1 & 17.8\erpm{0.4}{0.8} & 0.09 & 0.78 & 7.9\erpm{1.7}{1.3}\timestento{21} & 20.0\erpm{3.1}{3.1}&  2.18\\
            G287.3716+00.6444 & See Section \ref{g287} \\
            M8 E              & 13.5 & 15.8\erpm{2.8}{2.0} & 0.31 & 2.61 & 1.1\erpm{1.1}{7.9}\timestento{23} & 7.6\erpm{2.8}{2.0}& 1.38 \\
            G033.3891+00.1989 & 11.8 & 18.0\erpm{3.7}{2.8} & 0.24 & 2.05 & 9.4\erpm{6.6}{7.6}\timestento{21} & 19.9\erpm{16.1}{36.4}&  1.95 \\
            G308.9176+00.1231 & 42.6 & 29.0\erpm{3.9}{5.4} & 0.94 & 8.00 & 4.9\erpm{3.3}{12.9}\timestento{22} & 13.6\erpm{8.6}{15.0}&  1.14 \\
            G332.8256$-$00.5498 & 24.5 & 42.3\erpm{11.7}{4.4} & 0.59 & 5.08 & 2.2\erpm{\dag}{5.7}\timestento{21} & 18.6\erpm{17.6}{35.4}&  2.79 \\
            G347.0775$-$00.3927 & 17.7 & 30.6\erpm{14.3}{15.0} & 0.45 & 3.81 & 1.7\erpm{\dag}{7.4}\timestento{22} & 19.9\erpm{18.9}{86.2} &  0.62 \\
	    \hline
	  \end{tabular} 
	\end{center}
\small{$\ddag$: No uncertainties are presented as these inner radii
	are the inner radii of the region where CO emission is
	possible, i.e. $T\mathrm{<5000K}$. Therefore, decreasing the
	inner radii did not affect the quality of the fits.}\\
	\small{$\ast$: No uncertainties are presented as these outer
	radii are the outer radii of the region where CO emission is
	possible, i.e. $T\mathrm{>1000K}$. Therefore, increasing the
	outer radii did not affect the quality of the fits.}\\
	\small{$\dag$: $\chi^2$ flat down to
	$\mathrm{10^{18}cm^{-2}}$, the minimum surface number density
	considered. Therefore, no error is presented here.}
      \end{minipage}
    \end{center}

  \end{table*}
\end{center}

\subsection{Spectroastrometric signatures}

\label{pred_sig}

The spectroastrometric signatures associated with the spectra
presented in Fig. \ref{co_spec} are displayed in
Fig. \ref{spec_ast_obs}. The spectroastrometric data have been
re-binned by a factor of $\mathrm{\sim}$11 times the resolution
element. This rebinning factor was chosen as it was the maximum bin
size that could still resolve the expected spectroastrometric
signatures across the bandhead. The resulting average positional
precision is approximately 0.4~mas. In principle, this means we are
probing au size scales at the kpc distances of the sample. Here, we
compare the spectroastrometric data to the best-fitting models to
determine whether these data can be used to further probe the
circumstellar environments of the sample.

\smallskip

The spectroastrometric signatures associated with the best-fitting models
are also presented in Fig. \ref{spec_ast_obs}. The predicted
signatures are generally small, and to illustrate their appearance, an
enhanced spectrum is shown offset in each panel. The largest excursion
occurs on the blue side of the bandhead, which is due to the asymmetry
of the bandhead profile. Over the blue shoulder of the bandhead all
the emission is blue-shifted. Therefore, at these wavelengths the side
of the disc coming towards the observer is significantly brighter than
the other side, which results in the prominent spectroastrometric
signature at these wavelengths. Conversely, red-wards of the bandhead
peak the receding side of the disc is brightest, resulting in a
positional excursion in the opposite direction. However, at these
wavelengths red-shifted emission from the peak of the bandhead is
mixed with blue-shifted emission of ro-vibration lines at longer
wavelengths.  As a consequence, the spectroastrometric signature is
smaller to the red of the bandhead peak than on the blue side. Over
individual, resolved ro-vibrational lines the two sides of the disc
are almost equally bright, and thus the excursions at positive and
negative velocities are nearly symmetrical, resulting in the classic
`S' shaped signature of a disc \citep[][Wheelwright et al. in
prep.]{Pontoppidanetal2008,Vanderplas2009}. As the individual
ro-vibrational lines are located close together the adjacent
positional signatures result in an oscillating signature.

\smallskip

The largest predicted excursions are of the order 0.01 mas, which is
much smaller than the precision of the spectroastrometric data. The
small scale of the spectroastrometric signatures is largely due to the
large distances to the science objects, coupled with relatively weak
CO emission, which is typically 10 per cent of the continuum flux. The
largest predicted signatures are those of IRAS 08576$-$4334 and M8E
(both 0.017~mas), due to their proximity. Even with a positional
precision of 0.15 mas, the best in the sample, such excursions would
not be detected.

\smallskip

While the spectroastrometric data do not provide useful constraints to
the model fitting process, they are entirely consistent with the best-fitting models of emission originating in circumstellar discs. Moreover,
the data do provide additional constraints on the source of the
emission. We note that the spectroastrometric signature of IRAS
08576$-$4334 exhibits no detectable excursion with a precision of
approximately 0.7~mas. This is contrary to the finding of
\citet{Grave2007}, who report an excursion of approximately 18~mas,
corresponding to a minimum size of 40~au at 2.17~kpc, and likely
several times this. We suggest that the feature in their data is an
instrumental artefact, which is not present in our data as our
practice of subtracting anti-parallel position spectra eliminates
artefacts (the data of \citet{Grave2007} were obtained using a
constant PA.)

\section{Discussion}
\label{disc}

\subsection{A discussion on the best-fitting parameters}

The best-fitting inner disc radii are typically of the order 0.1~au,
and the best-fitting outer disc radii are generally 1--5~au. These
values are consistent with those of \citet{Bik2004}, and indicate that
the CO emission arises from small scale circumstellar discs. At such
small distances from the central star, CO molecules should be
disassociated by ultraviolet photons. However, the best-fitting number
surface densities are sufficient for the CO molecules to self-shield,
as found by \citet{Bik2004}. We note that the simplistic treatment of
the temperature gradient within the disc allows CO to be excited
within 1~au of the central star. However, a full treatment of viscous
heating in such a disc keeps the mid-plane temperature above
$\mathrm{\sim}$5000~K out to approximately 1~au
\citep{Vaidya2009}. This would have the effect of moving the inner
radius of the CO emitting region further from the central star. The
difference between the best-fitting inner radii and the radius at
which the mid-plane temperature of a massive accretion disc drops
below 5000K is of the order a few, and thus is unlikely to
significantly affect the best-fitting properties. {{In addition,
the viscous heating is dependent upon the accretion rate, which is
uncertain and thus incorporating viscous heating in the model would
introduce an additional unknown}}. Therefore, we judge the simplistic
temperature gradient to be sufficient for the purposes of this paper.

\smallskip

In the model the excitation temperature in the disc varies as a
function of radius. If the disc is sufficiently large the outer
temperature falls below 1000~K and the outer cells contribute no
flux to the total spectrum. In this situation increasing the outer
radius will not affect the resultant $\chi^2$, as the output spectrum
will not be changed. However, the physical disc may well be much
larger than the CO emitting region, which constitutes the warmer and
denser part of the disc. In this respect the outer regions of the CO
emitting region are consistent with the models of \citet{Vaidya2009},
who show that accretion discs around massive protostars may well be
stable out to approximately 80~au. The dust sublimation radii
associated with MYSOs are typically greater than 10~au
\citep{Dewit2007,Vaidya2009}. Therefore, the CO emission of the best-fitting models originates from within the dust sublimation radius, and
thus offers a unique tracer of gas prior to its accretion onto the
central star.

\smallskip

The inner radii set the initial excitation temperature, which is a
function of the distance from the central star. As a result, changing
the inner radius effects both the temperature in the disc and the
rotational broadening of the bandhead. To consistently determine the
temperature structure of the disc, and thus more stringent
constraints, a detailed radiative transfer model of the disc is
required. However, even if the gas disc were modelled in detail, the
properties of the central star are essentially unknown. For example
the star may swell to several times its main sequence size as a
consequence of high accretion rates \citep{Hosokawa2009}, which will
reduce its effective temperature. Therefore, even a sophisticated disc
model wold be subject to several unknowns.

 \smallskip

The best-fitting line-widths are generally greater than the value of
$\mathrm{\sim 4\,km\,s^{-1}}$ determined by \citet{Berthoud2007}, who
fit the CO $\mathrm{1^{st}}$ overtone emission of the Be star 51 Oph
with a circumstellar disc model. The line widths are also
approximately ten times the thermal broadening due to motion of the CO
molecules. Therefore, the dominant contribution must be due to
turbulence. However, we do not consider shear broadening in the
model. If this were taken into account, the turbulent velocity may
well be less than the best-fitting values presented in Table
\ref{fit}. The properties of accretion discs around MYSOs has only
recently begun to be examined in detail \citep{Vaidya2009}, and
incorporating vertical disc structure to the disc model \citep[which
is required to assess shear broadening][]{Horne1986,Hummel2000} is
beyond the scope of this paper. Therefore, we neglect shear
velocity. However, it is important to note that neglecting vertical
structure may artificially limit the extent of the CO emitting
region. In a disc with a vertical extent the upper regions of the disc
may be hot enough to excite CO emission at radii where the mid-plane
temperature has dropped below the required
$\mathrm{\sim2000~K}$. Therefore, incorporating vertical structure in
the model may allow CO emission from a wider range of radii than the
thin disc model. This would have the effect of increasing the
resulting spectro-astrometric signatures, and thus the predictions
presented here may be lower limits.

\smallskip

\subsection{Comparisons with previous work}

Little is known about the circumstellar environments of the RMS
objects. In particular, there are no previous high resolution studies
with which to compare the best-fitting model parameters. However, the two
non-RMS objects in the sample, M8 E and IRAS 08576$-$4334 have been
both been studied previously. Here, we assess whether the best-fitting
models for these objects are consistent with other
observations. 

\smallskip

\citet{Bik2004} determine the inclination of IRAS 08576$-$4334 to be
$\mathrm{27^{{\circ}+2}_{-14}}$ based on fitting the CO
$\mathrm{1^{st}}$ overtone bandhead. The best-fitting value of
$\mathrm{\sim18^{\circ}}$ determined here is thus consistent with the
previous value. This might be expected as we apply the same
methodology, none the less, this provides an important check that the
results are consistent with previous work. Turning to M8E,
\citet{Simon1985} postulate that this object possesses an edge on
disc; contrary to our best-fitting inclination of
$\mathrm{\sim16^{\circ}}$. However, \citet{Linz2009} find an
inclination $\mathrm{<30^{\circ}}$ is required to fit the SED of
M8E. These authors suggest that the interpretation of the lunar
occultation data of \citet{Simon1985} is complicated by the presence
of scattered light and outflow cavities. Given that the conclusion of
\citet{Linz2009} is based on the SED of M8 E from the visible to mid
infrared, in addition to the grid of models of \citet{Robitaille2007},
we suggest the inclination determined by \citet{Linz2009} is currently
the best estimate, which is in agreement with the best-fitting
value. Therefore, we conclude that the best-fitting models are
consistent with results in the literature, although only a few are
available, indicating that the best-fitting model parameters are
representative of the circumstellar environments of the sample.

\subsection{G287.3716+00.6444}
\label{g287}

In all but one case, the disc model not only provided a good fit to
the data (as measured with $\chi^2$), but was also consistent with the
observed flux densities, and where they existed, previous
observations. Therefore, it would appear that small scale discs are
present around the majority of the sample. However, it was not
possible to fit the CO $\mathrm{1^{st}}$ overtone bandhead of
G287.3716+00.6444 with the circumstellar disc model used to fit the
other profiles. As shown in Fig. \ref{co_spec} the disc model could
not fit observed the bandhead shoulder, nor the slope red-wards of the
peak. Furthermore, the bandhead could not be fit with emission from an
isothermal, non-rotating body of CO. Therefore, we are led to consider
alternative scenarios to explain the emission and its profile.

\smallskip 

Besides hot, dense discs there are other viable sources of CO
$\mathrm{1^{st}}$ overtone bandhead emission. One such scenario is a
dense, neutral wind. \citet{Chandler1995} were able to fit the CO
$\mathrm{1^{st}}$ overtone bandheads of several YSOs with models of
neutral winds, but note that the mass loss rates required were
relatively high, up to $\mathrm{10^{-6}\,M_{\odot}yr^{-1}}$. \citet{Chandler1995} consider it unlikely
CO $\mathrm{1^{st}}$ overtone emission originates in a wind as such
mass loss rates are much higher than observed for solar mass
YSOs. However, the winds of MYSOs may well lead to mass loss rates of
$\mathrm{10^{-6}\,M_{\odot}yr^{-1}}$ \citep{Drew1993}. Consequently, CO
$\mathrm{1^{st}}$ overtone emission from a dense wind should perhaps
be re-considered in this case. Alternatively, \citet{Scoville1983}
propose that the CO $\mathrm{1^{st}}$ overtone bandhead emission of
the Becklin-Neugebauer object is created in shocks (as the observed
velocity dispersion and estimated emitting area are both small).

 \smallskip

We note that \citet{Blum2004} also report that the CO
$\mathrm{1^{st}}$ overtone bandhead of one of their sample (of four
YSOs) was difficult to fit with a model of CO emission from a
circumstellar disc. Following the example of \citet{Kraus2000}, they
suggest that this may be due to the circumstellar disc exhibiting an
outer bulge, which shields the inner regions of the disc. The effect
of this would be to limit the visible CO emitting region to low
velocity regions, resulting in a narrow profile. In this case, however, it is
not the width of the profile we cannot fit, but rather the slope
of the profile red-wards of the bandhead peak. Most formation
scenarios, such as winds and discs, result in excess blue-shifted
emission, therefore this bandhead profile is difficult to explain.

\smallskip

It is conceivable that the emission from several, discrete regions in
a shock, will have different excitation temperatures, and when
superimposed, will result in a different slope to the bandhead than
the disc model. If such shocks exist, the shocked region must be located
close to the central star, within a few au, as we do not see a
positional excursion in the spectroastrometric data. However, it is
difficult to envisage a scenario in which the shock emission is
predominately red-shifted. As an alternative it may be that the
emission does originate in a disc, but the receding part of the disc
is significantly brighter than the approaching side, similar to the
V/R variations exhibited by Be stars
\citep[e.g.][]{Hanuschik1995}. Regardless, it is apparent that while
the majority of MYSO CO bandheads are well fit by models of
circumstellar discs, the circumstellar environments of MYSOs are 
not yet completely understood.

\section{Conclusions}
\label{Conclusions}

In this paper we present high resolution near infrared
spectroastrometry around 2.3$\mathrm{\mu m}$ of a sample of young
stellar objects, most of which are drawn from the RMS survey. The RMS
constitutes the largest catalogue of massive young stellar objects to
date, and thus provides a unique sample to study massive star
formation. We model the CO $\mathrm{1^{st}}$ overtone bandhead
emission detected with a simple model of a circumstellar disc in
Keplerian rotation. In addition, the sub milli-arcsecond precision
spectroastrometric data are compared to the spatial signatures of the
best-fitting models.

 \smallskip

The observed bandhead profiles are, on the whole, well fit by models
of the emission originating in circumstellar discs. This concurs with
the generally accepted view that this emission arises in small scale
accretion discs. The observed bandhead of one object cannot
be fit with a circumstellar disc model. This could be due to the
emission emanating from an asymmetric disc or shock. This highlights
that the circumstellar environments of massive young stellar
objects are still not completely understood.

\smallskip

No spatial signatures of discs are revealed in the spectroastrometric
data, which have a precision of approximately 0.4~mas. This is
entirely consistent with the sizes of the best-fitting models. Due to the
small sizes of the discs, the large distances to them and the decrease
in brightness with radius the predicted signatures are of the order
of $\mathrm{0.01mas}$. This is well below the current detection
limit. However, the predicted signatures could perhaps be revealed by
differential phase observations with AMBER. Currently, MYSOs are below
the sensitivity limit of AMBER, but PRIMA should allow us to observe
such objects and probe the spatial distribution of the CO emission at
an even higher precision.

\smallskip

To summarise, in general the model of emission from a circumstellar
disc provides a good fit to the observed bandheads, and is consistent
with the observed flux densities. This indicates that the majority of
MYSOs with CO $\mathrm{1^{st}}$ overtone emission possess small-scale,
circumstellar discs. In turn, this provides further evidence that
massive stars form via disc accretion, as suggested by the simulations
of \citet{Krumholz2009}.

\section*{Acknowledgements}

H.E.W gratefully acknowledges a PhD studentship from the Science and
Technology Facilities Council of the United Kingdom. R.D.O
thanks the Leverhulme Trust for the award of a research
fellowship. This paper is partly based on observations obtained at the
Gemini Observatory, which is operated by the Association of
Universities for Research in Astronomy, Inc., under a cooperative
agreement with the NSF on behalf of the Gemini partnership: the
National Science Foundation (United States), the Science and
Technology Facilities Council (United Kingdom), the National Research
Council (Canada), CONICYT (Chile), the Australian Research Council
(Australia), Minist\'{e}rio da Ci\^{e}ncia e Tecnologia (Brazil) and
Ministerio de Ciencia, Tecnolog\'{i}a e Innovaci\'{o}n Productiva
(Argentina). This paper is partly based on observations obtained with
the Phoenix infrared spectrograph, developed and operated by the
National Optical Astronomy Observatory. This paper has made use of
information from the Red MSX Source survey database at
www.ast.leeds.ac.uk/RMS which was constructed with support from the
Science and Technology Facilities Council of the UK.

\bibliographystyle{mn2e}
\bibliography{RMS_CO_paper}

\label{lastpage}
\end{document}